\documentclass[12pt,onesided]{amsart}

\usepackage{natbib}

\usepackage[default,oldstyle,scale=0.95]{opensans}
\usepackage[T1]{fontenc}
\usepackage{ae}

\usepackage{blindtext}

\usepackage{geometry}
\usepackage{parallel}
\usepackage{parcolumns}
\usepackage{fancyhdr}
\usepackage[export]{adjustbox}
\usepackage{etoolbox}

\usepackage{array}
\usepackage{amsmath}
\usepackage{amssymb}
\usepackage{amsthm}
\usepackage{amsfonts}
\usepackage{relsize}
\usepackage{mathtools}
\usepackage{bm}
\usepackage{siunitx}

\usepackage[section]{placeins}

\usepackage{tabularx}
\usepackage{booktabs}
\usepackage{multicol}
\usepackage{multirow}
\usepackage{longtable}

\usepackage[font={small},labelfont=bf]{caption}

\usepackage[bottom]{footmisc}

\usepackage{enumerate}

\usepackage{paralist}

\usepackage[pdftex,hyperref,x11names]{xcolor}

\usepackage[pdftex]{hyperref}

\usepackage{subfig}
\usepackage{graphicx}
\usepackage[space]{grffile} 
\usepackage{placeins} 
\usepackage{tikz}

\usepackage{rotating}
\usepackage{lscape}

\usepackage[doublespacing]{setspace}

\hypersetup{%
	unicode=false,          
	pdftoolbar=true,        
	pdfmenubar=true,        
	pdffitwindow=false,     
	pdfstartview={FitH},    
	pdfnewwindow=true,%
	pdfauthor={Shahryar Minhas},%
	pdftitle={Decomposing Network Influence: Social Influence Regression},%
	colorlinks,%
	citecolor=black,%
	filecolor=black,%
	linkcolor=black,%
	urlcolor=RoyalBlue4
}

\title{Decomposing Network Influence: Social Influence Regression}
\vspace{\baselineskip}

\author[Minhas]{Shahryar Minhas}
\address{Shahryar Minhas: Department of Political Science}
\curraddr{Michigan State University}
\email[Corresponding author]{minhassh@msu.edu}

\author[Hoff]{Peter D. Hoff}
\address{Peter D. Hoff: Department of Statistics}
\curraddr{Duke University}
\email{peter.hoff@duke.edu}

\date{\today}

\makeatletter

\newcommand{\pkg}[1]{{\fontseries{b}\selectfont #1}}

\definecolor{red1}{RGB}{253,219,199}
\definecolor{red2}{RGB}{244,165,130}
\definecolor{red3}{RGB}{178,24,43}

\definecolor{green1}{RGB}{229,245,224}
\definecolor{green2}{RGB}{161,217,155}
\definecolor{green3}{RGB}{49,163,84}

\definecolor{blue0}{RGB}{255,247,251}
\definecolor{blue1}{RGB}{222,235,247}
\definecolor{blue2}{RGB}{158,202,225}
\definecolor{blue3}{RGB}{49,130,189}
\definecolor{blue4}{RGB}{4,90,141}

\definecolor{purple1}{RGB}{191,211,230}
\definecolor{purple2}{RGB}{140,150,198}
\definecolor{purple3}{RGB}{140,107,177}

\definecolor{brown1}{RGB}{246,232,195}
\definecolor{brown2}{RGB}{223,194,125}
\definecolor{brown3}{RGB}{191,129,45}

\let\bbordermatrix\bordermatrix
\patchcmd{\bbordermatrix}{8.75}{4.75}{}{}
\patchcmd{\bbordermatrix}{\left(}{\left[}{}{}
\patchcmd{\bbordermatrix}{\right)}{\right]}{}{}

\begin{document}
\maketitle\thispagestyle{empty}

\small{\singlespacing{
\begin{abstract}
Understanding network influence and its determinants are key challenges in political science and network analysis. Traditional latent variable models position actors within a social space based on network dependencies but often do not elucidate the underlying factors driving these interactions. To overcome this limitation, we propose the Social Influence Regression (SIR) model, an extension of vector autoregression tailored for relational data that incorporates exogenous covariates into the estimation of influence patterns. The SIR model captures influence dynamics via a pair of $n \times n$ matrices that quantify how the actions of one actor affect the future actions of another. This framework not only provides a statistical mechanism for explaining actor influence based on observable traits but also improves computational efficiency through an iterative block coordinate descent method. We showcase the SIR model's capabilities by applying it to monthly conflict events between countries, using data from the Integrated Crisis Early Warning System (ICEWS). Our findings demonstrate the SIR model's ability to elucidate complex influence patterns within networks by linking them to specific covariates. This paper's main contributions are: (1) introducing a model that explains third-order dependencies through exogenous covariates and (2) offering an efficient estimation approach that scales effectively with large, complex networks.
\end{abstract}
}}

\newpage\setcounter{page}{1}

\section*{\textbf{Motivation}}

Network influence shapes political outcomes across diverse domains, yet precisely measuring and explaining these patterns of influence remains a challenge. We define network influence broadly as the impact that one actor's actions or decisions have on the behavior of others within a network, whether through direct or indirect connections. Numerous studies have demonstrated the critical role of network influences in explaining a wide range of political phenomena, from subnational policy diffusion to interstate conflict dynamics \citep{cranmer:etal:2015a, beardsley:etal:2020, nieman:etal:2021, edgerton:2024}. A prominent approach to measuring influence relies on latent variable models, which position actors in a social space based on mechanisms such as transitivity and/or stochastic equivalence \citep{gade:etal:2019, huhe:etal:2021, edgerton:2024}. However, while these models can effectively describe the overall structure of a network, they frequently fall short in providing detailed explanations for the specific influence that actors exert on one another. This limitation arises because these models typically attribute influence to broad network features, without accounting for the exogenous factors that might drive such influence, leaving the mechanisms behind these interactions underexplored and poorly understood.

To address this limitation, we propose a novel approach: the Social Influence Regression (SIR) model. This model extends vector autoregression techniques to relational data, allowing us to capture the influence of one actor on another over time while simultaneously incorporating exogenous covariates. The SIR model operates by estimating a pair of $n \times n$ matrices that measure sender- and receiver-level influence patterns, taking the form:

\[ y_{ij,t} = \sum a_{ii'} b_{jj'} x_{i'j',t-1} + e_{ij} \]

where $a_{i,i'}$ captures how the previous actions of $i'$ affect those of $i$, and $b_{j,j'}$ indicates how actions toward target $j$ are influenced by prior actions toward $j'$.

The key innovation of the SIR model is its ability to explicitly account for the role of exogenous covariates in shaping these influence patterns. While traditional models in this vein are effective at uncovering patterns within the network, they often fall short of explaining these patterns in terms of observable actor-level or dyad-level attributes. The SIR model bridges this gap by linking actor positioning in the latent social space directly to exogenous covariates, thus providing a more interpretable and theoretically grounded understanding of network dynamics.

We apply this approach to data from the Integrated Crisis Early Warning System (ICEWS) event data project. Using the SIR model, we estimate the extent to which actors within the material conflict network influence one another and, crucially, explore how characteristics such as alliances or economic relationships explain the observed influence patterns. Our findings demonstrate that the SIR model significantly improves out-of-sample performance compared to existing methods. This improvement underscores the model's effectiveness and offers new insights into the drivers of influence in international relations. By providing a more precise and interpretable representation of network dynamics, this work advances both the methodology of network analysis and the substantive understanding of international conflict processes.

The rest of the paper proceeds as follows. In Section 2, we introduce the model in detail, describing its theoretical foundations and estimation procedure. Section 3 presents our empirical application to the ICEWS data, including a description of the data, model specification, and results. We pay particular attention to how alliance relationships and trade flows influence conflict behavior. Section 4 provides a performance comparison of the SIR model against alternative approaches, demonstrating its superior out-of-sample predictive power. Finally, Section 5 concludes with a discussion of the implications of our findings for international relations theory and suggestions for future research directions in network analysis within political science.

\section*{\textbf{Methods}}

\subsection*{Bilinear network autoregression model}

The bilinear network autoregression model provides a framework for understanding how interactions between actors in a network can be modeled over time \citep{minhas:etal:2016}. In this paper, we extend this framework by presenting the Social Influence Regression (SIR) model, which not only offers a novel method to explain the factors driving the influence parameters in the bilinear autoregression model but also introduces a more efficient estimation scheme. Our iterative block coordinate descent method dramatically accelerates the estimation process compared to the Bayesian approach originally used in the bilinear autoregression framework, making it much faster and more scalable for large networks.

Many studies examine the flows or linkages among actors, such as whether two countries are in conflict with one another. These interactions are often represented as a matrix, as shown in Figure~\ref{fig:matrix}. This matrix is $n \times n$, where $n$ denotes the number of actors in the network. The off-diagonal elements $y_{ij}$ represent the interaction between actors $i$ and $j$. In undirected data, $y_{ij}$ may indicate, for example, whether $i$ and $j$ are allied. In directed data, the rows represent senders and the columns represent receivers, so $y_{ij}$ would indicate an action sent from $i$ to $j$. The diagonal elements are typically undefined, indicating that actors do not interact with themselves.

\begin{figure}[ht]
\centering
\begin{minipage}{.45\textwidth}
	\begin{equation*}
	Y_{t} = \bbordermatrix{
		~ & i  & \ldots & j & \ldots & k \cr
		i & NA  & \ldots & y_{ji} & \ldots & y_{ik} \cr
		\vdots & \vdots & \ddots & \vdots & \ddots & \vdots  \cr
		j & y_{ji}  & \ldots & NA  & \ldots & y_{jk} \cr
		\vdots & \vdots & \ddots & \vdots & \ddots & \vdots \cr
		k & y_{ki}  & \ldots & y_{kj}  & \ldots & NA \cr
		}
	\end{equation*}
	\caption{Matrix representation of a dyadic, relational measure for one time point.}
	\label{fig:matrix}
\end{minipage}
\begin{minipage}{0.45\textwidth}
	\centering
	\resizebox{.7\textwidth}{!}{\begin{tikzpicture}



	
	\begin{scope}[xshift=4.5cm, yshift=1cm]
	\node{
		\begin{tikzpicture}[scale=.5]
			 \draw[thin, black,fill=blue4] (0,0) grid (4,4) rectangle (0,0) ;
		\end{tikzpicture}
	};
	\end{scope}

	\begin{scope}[xshift=4cm, yshift=.5cm]
	\node[](blue){
		\begin{tikzpicture}[scale=.5]
			 \draw[thin, black,fill=blue3] (0,0) grid (4,4) rectangle (0,0) ;
		\end{tikzpicture}
	};
	\end{scope}

	\begin{scope}[xshift=3.5cm]
	\node{
		\begin{tikzpicture}[scale=.5]
			 \draw[thin, black,fill=blue2] (0,0) grid (4,4) rectangle (0,0) ;
		\end{tikzpicture}
	};
	\end{scope}
	
\end{tikzpicture}}
	\caption{Array representation of a longitudinal dyadic measure. Darker shading indicates later time periods.}
	\label{fig:tensViz}
\end{minipage}
\end{figure}

Figure~\ref{fig:matrix} captures interactions between actors at a single point in time. However, interactions are often observed over a series of time points. To represent longitudinal network data, we stack these adjacency matrices into an array, as shown in Figure~\ref{fig:tensViz}. Specifically, let $Y = \{Y_t : t = 1, \ldots, T\}$ be a time series of relational data, where $T$ represents the number of time points. The resulting array has dimensions $n \times n \times T$. The bilinear autoregression model is designed to estimate dependencies in such structures by regressing the network at one time point on its lag. The relationship between these time points is captured by a pair of matrices that reflect sender and receiver dependence patterns for each dyad.

A generalized bilinear autoregression model for $Y$ is given by:

\begin{align*}
	E [ y_{i,j,t} ] &= g( \mu_{i,j,t} ), \\
	\{ \mu_{i,j,t} \} &= M_t = A X_t B^{\top}, \\
	\{ \mu_{i,j,t} \} & = a_i^{\top} X_{t} b_j,
\end{align*}

where $x_{i,j,t}$ is a function of $y_{i,j,t}$, such as $ \tilde{x}_{i,j,t} \sim \log(y_{i,j,t-1} + 1)$.

In the application, we explore an example involving count data, where $Y$ is a time series of matrices defining count-based events between actors. For instance, we model $ y_{i,j,t} \sim \text{Poisson}(e^{\mu_{i,j,t}})$, with $ \tilde{x}_{i,j,t} = \log(y_{i,j,t-1} + 1)$. This framework is extendable to other distributions, as it is based on a generalized bilinear model. The matrices $A$ and $B$ are $n \times n$ "influence parameters." The value of $a_{ii'}$ captures how predictive the actions of country $i'$ at time $t-1$ are of the actions of country $i$ at time $t$, while the value of $b_{jj'}$ captures how predictive the actions directed at country $j'$ at time $t-1$ are of the actions directed towards country $j$ at time $t$.

For example, in a bilinear autoregression model on conflict involving the United Kingdom (GBR) and the United States of America (USA), if $ a_{\text{GBR}, \text{USA}} $ is greater than zero, it implies that countries with which the USA initiated or continued a conflict in period $t-1$ are likely to also face conflict from GBR in period $t$. This suggests that GBR's future actions are influenced by the USA, or more concretely, the USA's actions are predictive of GBR's.

While the bilinear autoregression model provides a robust framework for capturing dependence patterns within network data, it falls short in its ability to explain the underlying mechanisms driving these influence patterns. Specifically, the model does not incorporate exogenous factors that may account for why certain actors exert influence within the network, limiting its interpretability and theoretical grounding. The Social Influence Regression (SIR) model, introduced in the next section, addresses this gap by incorporating exogenous covariates, offering a more detailed and interpretable understanding of what drives the influence parameters, $a$ and $b$, within the network. This innovation not only enhances the explanatory power of the model but also enables greather theoretical insight into the dynamics of network interactions.

\subsection*{Social influence regression}

The SIR model explains influence in terms of covariates. Particularly, to determine the characteristics of $i$ or $i'$ that are related to the influence $a_{ii'}$, we consider a linear regression model for $a_{ii'}$ and $b_{jj'}$, given by $a_{ii'} = \alpha^{\top} w_{ii'}$ and $b_{jj'} = \beta^{\top} w_{jj'}$, where $w_{ii'}$ is a vector of nodal and dyadic covariates specific to pair $ii'$ that we are using to estimate influence. The application we present in the following section has time-varying covariates, which this model is able to account for through time varying influence parameters: $a_{ii't} = \alpha^{\top} w_{ii't}$ and $b_{jj't} = \beta^{\top} w_{jj't}$.

The network autoregression model can be expressed as:

\begin{align*}
	\mu_{i,j,t}  & = \sum_{i'j'} a_{ii't}  x_{i'j't} b_{jj't} = \sum_{i'j'} \alpha^{\top}w_{ii't}  x_{i'j't} w_{jj't}^{\top} \beta \\
	&= \alpha^{\top} \left (\sum_{i'j'} x_{i'j't} w_{ii't} w_{jj't}^{\top} \right ) \beta = \alpha^{\top} \tilde{X}_{ijt} \beta \\
\end{align*}

Typically, $y_{i,j,t}$ also has covariates. For example, we might want to condition estimation of the parameters on a lagged version of the dependent variable, $y_{i,j,t-1}$, a measure of reciprocity, $y_{j,i,t-1}$, and other exogenous variables. In the case of estimating a model on material conflict between a pair of countries, this might include other exogenous aspects such as the geographical distance between a pair of countries. These additional exogeneous parameters can be accommodated with a model of the form:

\begin{align*}
	\mu_{i,j,t} = \theta^{\top} z_{i,j,t} +  \alpha^{\top} \tilde{X}_{ijt} \beta,
\end{align*}

where $z_{i,j,t}$ represents the design array incorporating parameters that may have a direct effect on the dependent variable. The model presented here is a type of low-rank matrix regression, where we are regressing the outcome $y_{ij,t}$ on the matrix $X_{ij,t}$. An unconstrained (linear) regression would be expressed as $\mu_{ij,t} = \theta^{\top} z_{ij,t} +  \langle C,  {X}_{ij,t} \rangle$, where $C$ is an arbitrary $p\times p$ matrix of regression coefficients to be estimated. In contrast, the regression specified above restricts $C$ to be rank one, that is, expressible as $C=\alpha \beta^{\top}$. This follows from the identity that $\langle \alpha \beta^{\top}, {X}_{ij,t} \rangle  = \alpha^{\top} {X}_{ij,t} \beta$. Low rank matrix regression models have been considered by \citet{li:etal:2010} and \citet{zhou:etal:2013}.

\subsection*{Estimation}

The estimation of the parameters $\{\theta, \alpha, \beta\}$ in the bilinear network autoregression model is challenging due to the bilinear nature of the model. To address this, we employ an iterative block coordinate descent method, which alternately optimizes the parameters by treating one set of parameters as fixed while optimizing over the others. Specifically, the estimation process capitalizes on the fact that for fixed $\beta$, the model is linear in $\theta$ and $\alpha$, and for fixed $\alpha$, it is linear in $\theta$ and $\beta$. The conditional likelihood function for each of these cases can thus be optimized using standard techniques for generalized linear models (GLMs), specifically through iterative weighted least squares (IWLS).

The model can be expressed as follows:

\begin{align*}
	\mu_{i,j,t} & = ( \theta^{\top} \ \alpha^{\top} )
	\left ( \begin{array}{c} z_{i,j,t} \\ \tilde{X}_{ijt} \beta \end{array} \right ) \\
	 &= ( \theta^{\top} \ \beta^{\top} )
	\left ( \begin{array}{c} z_{i,j,t} \\ \tilde{X}_{ijt}^{\top} \alpha \end{array} \right )
\end{align*}

where $\mu_{i,j,t}$ represents the expected value of the interaction between actors $i$ and $j$ at time $t$, $z_{i,j,t}$ denotes the vector of covariates associated with the dyad $(i, j)$ at time $t$, and $\tilde{X}_{ijt}$ is the matrix of explanatory variables.

Given this setup, the parameters $\theta$, $\alpha$, and $\beta$ are estimated through the following iterative block coordinate descent procedure:

\begin{enumerate}
    \item \textbf{Initialize}: Start with an initial guess for $\beta$.
    \item \textbf{Iterate}: 
    \begin{enumerate}
        \item \textbf{Optimize $\theta$ and $\alpha$}: With $\beta$ fixed, estimate $\theta$ and $\alpha$ by maximizing the conditional log-likelihood function using iterative weighted least squares (IWLS). This is equivalent to fitting a generalized linear model (GLM) with the response variable $y$ regressed on the design matrix $\tilde{X}$, which is constructed by concatenating $z_{ij,t}$ and $X_{ij,t} \beta$ for each dyad $(i, j)$ and time $t$.
        \item \textbf{Optimize $\theta$ and $\beta$}: With $\alpha$ fixed, estimate $\theta$ and $\beta$ by maximizing the conditional log-likelihood function using IWLS. This step is analogous to step (a) but with the design matrix $\tilde{X}$ constructed by concatenating $z_{ij,t}$ and $X_{ij,t}^{\top} \alpha$.
    \end{enumerate}
    \item \textbf{Convergence}: Repeat steps (a) and (b) until the parameters converge, i.e., until the changes in the parameter estimates between iterations fall below a specified tolerance level.
\end{enumerate}

The iterative process leverages the fact that for each subproblem, the estimation reduces to a series of low-dimensional GLM optimizations. By solving these smaller problems iteratively, the overall estimation problem, which is bilinear and thus non-trivial, becomes manageable.

To implement the above steps, consider the following detailed process for step 1 (optimizing $\theta$ and $\alpha$):

\begin{enumerate}
    \item \textbf{Construct the Design Matrix}:
    \begin{enumerate}
        \item Let $\tilde{x}_{ij,t}$ be a vector of length $p+q$ obtained by concatenating $z_{ij,t}$ and $X_{ij,t} \beta$. Here, $p$ is the length of the vector $w_{ii'}$, and $q$ is the length of $z_{ij,t}$.
        \item Construct the design matrix $\tilde{X}$ with dimensions $n \times (n-1) \times T$ by $p+q$, where each row corresponds to a specific dyad $(i, j)$ at time $t$ and is equal to $\tilde{x}_{ij,t}$.
    \end{enumerate}
    \item \textbf{Create the Response Vector}:
    \begin{enumerate}
        \item Let $y$ be a vector of length $n \times (n-1) \times T$ consisting of the entries of $Y = \{ Y_1, \ldots, Y_T \}$, ordered to correspond with the rows of $\tilde{X}$.
    \end{enumerate}
    \item \textbf{Perform Poisson Regression}:
    \begin{enumerate}
        \item Obtain the MLEs for the Poisson regression of $y$ on $\tilde{X}$, which yields the conditional estimates of $\theta$ and $\alpha$.
    \end{enumerate}
    \item \textbf{Repeat for $\theta$ and $\beta$}:
    \begin{enumerate}
        \item In step 2 of the iterative algorithm, repeat the process by constructing the design matrix $\tilde{X}$ with $X_{ij,t}^{\top} \alpha$ replacing $X_{ij,t} \beta$ in step 1(a).
    \end{enumerate}
\end{enumerate}

The block coordinate descent method is particularly suited to this estimation problem because it effectively decomposes a high-dimensional optimization problem into more manageable subproblems. Each iteration refines the parameter estimates by focusing on a lower-dimensional subset of the parameters, thereby reducing the complexity of the problem.

Convergence is assessed by monitoring the change in parameter estimates across iterations. The process is typically stopped when the relative change in the log-likelihood or the parameter estimates between successive iterations falls below a predetermined threshold.\footnote{The choice of convergence criteria and initial values for $\beta$ can impact the speed and stability of convergence, and these factors are considered in the implementation of the algorithm. We will provide software in $\sf{R}$ to implement these types of models. The software will be hosted on CRAN and/or the corresponding author's github.}

The iterative nature of the estimation process ensures that the resulting parameter estimates are as close as possible to the true maximum likelihood estimates, given the bilinear structure of the model. This method provides a robust and efficient means of estimating the parameters in complex network models where direct optimization would be infeasible.

\subsection*{Inference}

After estimating the parameters of the SIR model, we assess the precision of these estimates by computing the standard errors. These standard errors are derived from the variance-covariance matrix, which is obtained from the inverse of the Hessian matrix of the log-likelihood function. Computing the standard errors, however, requires a consideration of the identifiability of the multiplicative parameters $ \alpha $ and $ \beta $.

The log-likelihood function for the SIR model, assuming a Poisson distribution for the count data, is given by:

\[
\ell(\theta, \alpha, \beta) = \sum_{i \neq j} \left( y_{ij,t} \log(\mu_{ij,t}) - \mu_{ij,t} \right),
\]

where $ \mu_{ij,t} = \exp(\eta_{ij,t}) $ and $ \eta_{ij,t} = Z_{ij,t}^\top \theta + \alpha^\top X_{ij,t} \beta $ represents the linear predictor incorporating both the fixed effects $Z_{ij,t}$ and the multiplicative terms involving the influence parameters $\alpha$ and $\beta$.

The parameters $ \alpha $ and $ \beta $ are not inherently identifiable because the term $ \alpha^\top X \beta $ is equivalent to $ \left(\frac{\alpha}{c}\right)^\top X \left(c \beta\right) $ for any scalar $ c $. To obtain meaningful derivative-based standard errors, we must first establish an identifiable parameterization of the model. This can be achieved by imposing a scale restriction on either $ \alpha $ or $ \beta $, or by fixing one element of either vector. The identifiable parameterization employed here restricts the first element of $ \alpha $ to be one, ensuring that the model parameters are uniquely determined.

The Hessian matrix $ H(\theta, \alpha, \beta) $ is composed of the second-order partial derivatives of the log-likelihood function with respect to the parameters $ \theta $, $ \alpha $, and $ \beta $, considering the identifiable parameterization:

\[
H(\theta, \alpha, \beta) = -\frac{\partial^2 \ell(\theta, \alpha, \beta)}{\partial \psi^2},
\]

where $\psi = (\theta, \alpha_{-1}, \beta)$ represents the identifiable parameter vector, excluding the fixed element of $\alpha$. For the identifiable parameterization, the Hessian matrix specifically consists of:

\[
H(\theta, \alpha, \beta) = \begin{pmatrix}
H_{\theta \theta} & H_{\theta \alpha_{-1}} & H_{\theta \beta} \\
H_{\alpha_{-1} \theta} & H_{\alpha_{-1} \alpha_{-1}} & H_{\alpha_{-1} \beta} \\
H_{\beta \theta} & H_{\beta \alpha_{-1}} & H_{\beta \beta}
\end{pmatrix},
\]

where each block is defined by the second derivatives:

\[
H_{\theta_k \theta_l} = -\sum_{i \neq j} \left( \mu_{ij,t} Z_{ij,k} Z_{ij,l} \right),
\]

\[
H_{\alpha_{ii'} \alpha_{ii''}} = -\sum_{i \neq j} \left( \mu_{ij,t} X_{i'j,t-1} \beta_{jj'} X_{i''j,t-1} \beta_{jj'} \right),
\]

\[
H_{\beta_{jj'} \beta_{jj''}} = -\sum_{i \neq j} \left( \mu_{ij,t} \alpha_{ii'} X_{i'j,t-1} \alpha_{ii'} X_{i'j'',t-1} \right).
\]

The mixed partial derivatives are:

\[
H_{\theta_k \alpha_{ii'}} = -\sum_{i \neq j} \left( \mu_{ij,t} Z_{ij,k} X_{i'j,t-1} \beta_{jj'} \right),
\]

\[
H_{\theta_k \beta_{jj'}} = -\sum_{i \neq j} \left( \mu_{ij,t} Z_{ij,k} \alpha_{ii'} X_{i'j,t-1} \right),
\]

\[
H_{\alpha_{ii'} \beta_{jj'}} = -\sum_{i \neq j} \left( \mu_{ij,t} X_{i'j,t-1} X_{i'j,t-1} \right).
\]

These expressions involve the expected counts $\mu_{ij,t} = \exp(\eta_{ij,t})$, which depend on the current estimates of $\theta$, $\alpha$, and $\beta$, and the design matrices $Z$ and $X$.

Given the identifiable parameterization, the standard errors for the parameters are derived from the inverse of the Hessian matrix:

\[
\text{Cov}(\hat{\theta}, \hat{\alpha}, \hat{\beta}) = -H^{-1}(\hat{\theta}, \hat{\alpha}, \hat{\beta}),
\]

where the standard errors are the square roots of the diagonal elements of this variance-covariance matrix:

\[
\text{SE}(\hat{\theta}_k) = \sqrt{ \left[ \text{Cov}(\hat{\theta}, \hat{\alpha}, \hat{\beta}) \right]_{kk} }.
\]

To obtain model-robust standard errors that are less sensitive to model misspecification, we employ a sandwich variance estimate:

\[
\widehat{\text{Var}}(\hat{\psi}) = H^{-1} S H^{-1},
\]

where $ S $ is the empirical information matrix, computed as:

\[
S = \sum_{i,j,t} \left( y_{i,j,t} - \mu_{i,j,t} \right) \dot{L}_{ij,t} \dot{L}_{ij,t}^\top,
\]

and $ \dot{L}_{ij,t} $ denotes the gradient (derivative) of the log-likelihood with respect to the parameters for a single observation $ y_{ij,t} $. The robust standard errors are then given by:

\[
\text{Robust SE}(\hat{\psi}_k) = \sqrt{ \left[ \widehat{\text{Var}}(\hat{\psi}) \right]_{kk} }.
\]

This robust variance-covariance matrix accounts for the variability in the score functions across observations, providing standard errors that are valid under a broader set of conditions.

Approximate standard errors and confidence intervals for the parameters are obtained from the derivatives of the log-likelihood function at the MLE. The asymptotic validity of these standard errors relies upon the assumption that the model is correctly specified. However, by using robust standard errors derived from the sandwich estimator, we ensure that the inferences drawn from the model are reliable even in the presence of potential model violations. In the application that follows in the next section, we utilize model-robust standard errors.

\begin{figure}[ht]
	\includegraphics[width=1\textwidth]{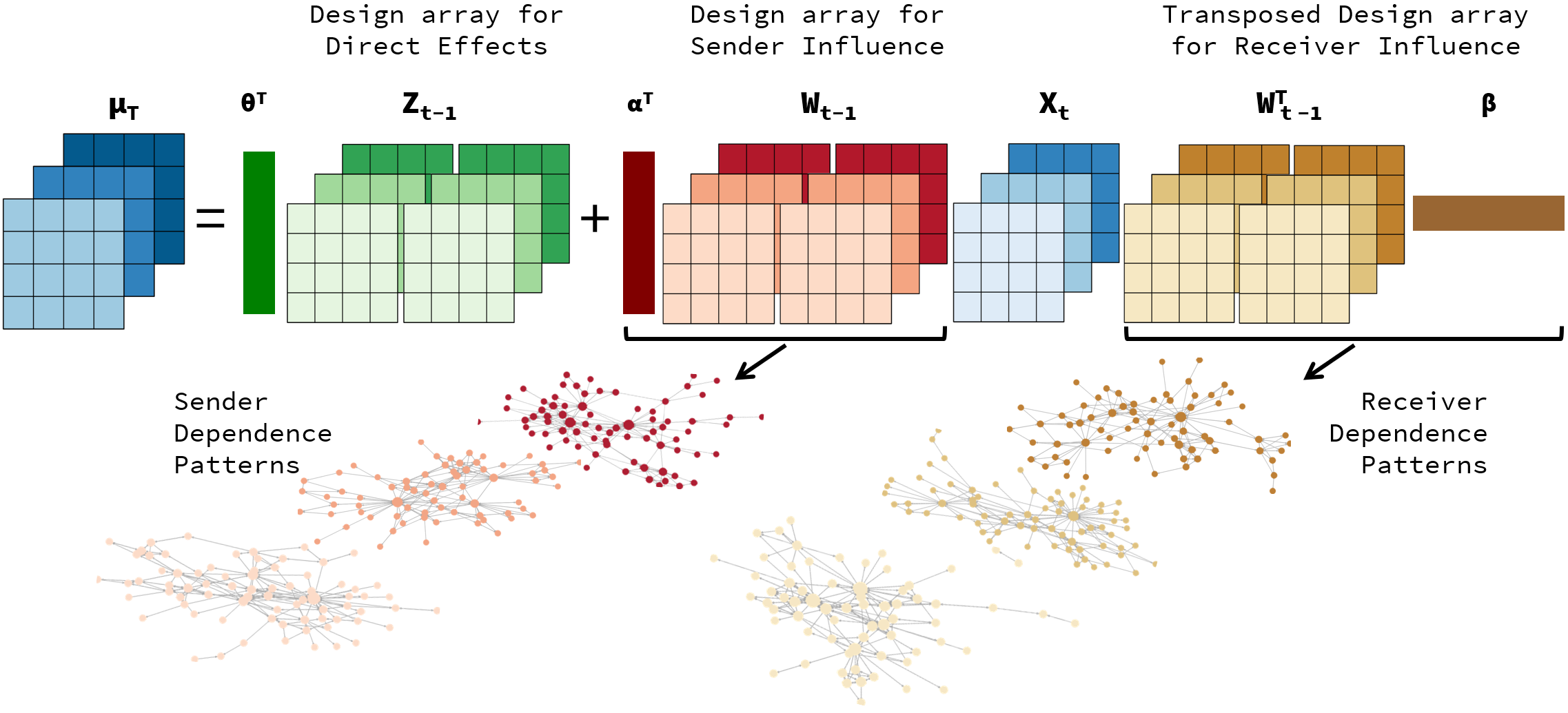}
	\caption{Visual summary of social influence regression model.}
	\label{fig:socialInfluenceViz}
\end{figure}

Figure~\ref{fig:socialInfluenceViz} provides a visual summary of this model. The array in the far left represents the network being modeled, the design array in green represents explanatory variables used to directly model linkages between dyads, and the $\theta$ vector includes the estimates of the effect those variables have on the network. To capture dependence patterns, a logged and lagged version of dependent variable are included, along with a design array containing a set of influence covariates, $W$; $\alpha$ and $\beta$ are vectors that capture parameter estimates for the effects of those influence covariates. A benefit of this framework is that once  estimated, linear combinations of the influence regression parameters permit visualizing the resulting sender and receiver dependence patterns in the network.

\section*{\textbf{Empirical Application}}

\subsection*{ICEWS Material Conflict}

A number of projects have arisen seeking to create large data sets of dyadic events through the automatic extraction of information from on-line news archives. This has made it empirically easier to study interactions among countries, as well as among actors such as NGOs within countries.

The two most well-known developments include the ICEWS event data project \citep{icews:2015:data} and the Phoenix pipeline \citep{oeda:2016}. For the purposes of this project we focus on utilizing the ICEWS database as it extends back farther in time. ICEWS draws from over 300 different international and national focused publishers \citep{boschee:etal:2015}. The ICEWS event data are based on a continuous monitoring of over 250 news sources and other open source material covering 177 countries worldwide. ICEWS consists of several components, including a database of over 38 million multilingual news stories going back to 1990 and present to last week. The ICEWS data along with extensive documentation have been made publicly available (with a one year embargo) on \url{dataverse.org} \citep{icews:2015:aggregations,icews:2015:data}. To classify news stories into socio-political topics, ICEWS relies on a augmented and expanded version of the CAMEO coding scheme \citep{schrodt:etal:2009}. The dictionaries, aggregations, ground truth data, and actor and verb dictionaries are publicly  available with a one year lag at the ICEWS data repository \url{https://dataverse.harvard.edu/dataverse/icews}. In addition,  the event coder has been made available publicly by the Office of the Director of National Intelligence.\footnote{Details at \url{http://bit.ly/2nS4nBU}.} This event coder, known as ACCENT, searches for the following information: a sender, a receiver, an action type, and a time stamp. The set of action types covered include activities between dyads such as ``Occupy territory'', ``Use conventional force'', and ``Impose embargo, boycott, or sanctions''. Then, the ontology provides rules through which the parsed story is coded. An example of a coded news story fitting this last category is:

\begin{quote}
	``\textit{President Bill Clinton has imposed sanctions on the Taliban religious faction that controls Afghanistan for its support of suspected terrorist Osama bin Laden, the White House said Tuesday.}''
\end{quote}

In this example, the actor designated as sending the action is the United States and the actor receiving it is Afghanistan. Dyadic measurements such as these are available for 249 countries, and the dataset is updated regularly. Currently, data up until March 2016 has been made publicly available on the ICEWS dataverse.

Our sample for this analysis focuses on monthly level interactions between countries in the international system from 2005 to 2012.\footnote{The ICEWS data extends to 2016 but we end at 2012 due to temporal coverage constraints among other covariates that we have incorporated into the model.} To measure conflict from this database we focus on what is often referred to as the ``material conflict'' variable. This variable is taken from the ``quad variable'' framework developed by \citet{duval:thompson:1980}. \citet{schrodt:yonamine:2013} defines the type of events that get drawn into this category as those involving, ``Physical acts of a conflictual nature, including armed attacks, destruction of property, assassination, etc.''.

\newcolumntype{M}[1]{>{\centering\arraybackslash}m{#1}}
\begin{figure}[ht]
	\centering
	\hspace{-80mm}
	\begin{tabular}{M{1cm}M{2.5cm}}
	    \scshape{January 2005} & \includegraphics[width=.8\textwidth]{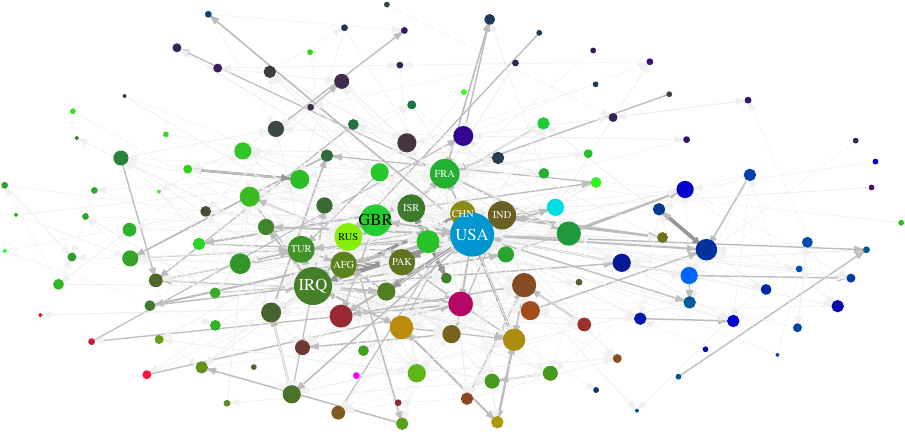}  \\
	    \scshape{December 2012} & \includegraphics[width=.8\textwidth]{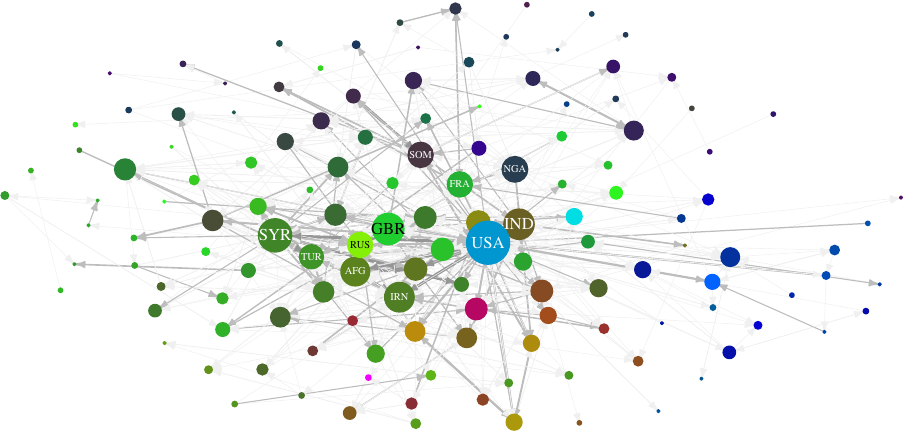} \\
	    \multicolumn{2}{c}{\includegraphics[width=.4\textwidth]{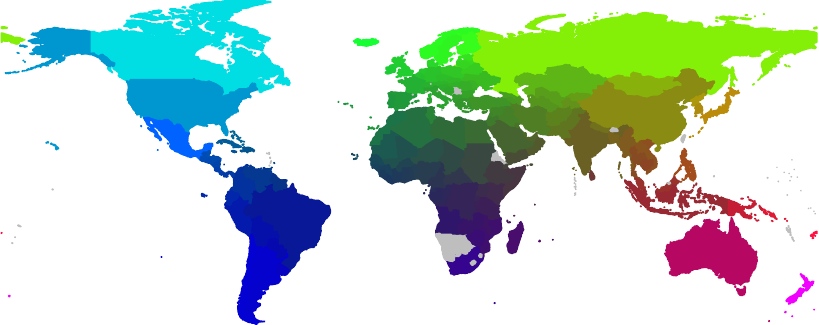}} \\
	\end{tabular}
	\caption{Network depiction of ICEWS Material Conflict events for January 2005 (top) and December 2012 (bottom).}
	\label{fig:icews}
\end{figure}
\FloatBarrier

Figure~\ref{fig:icews} visualizes the material conflict variable as a network, specifically, we provide snapshots of events between dyads along this relational dimension in January 2005 and December 2012. The size of the nodes correspond to how active countries are in the network, and each node is colored by its geographic position. An edge between two nodes designates that at least one material conflict event has taken place between that dyad, and arrows indicate the sender and receiver. Thicker edges indicate a greater count of material conflict events between a dyad.

In both snapshots, the United States is highly involved in conflict events occurring in the system both in 2005 and 2012. Additionally, other major powers such as Russia and the Great Britain are also frequently involved. Some notable changes are visible in the network. While in 2005 Iraq was highly involved in material conflict events by 2012 Syria became more active. Last, there is a significant amount of clustering by geography in this network. Conflict involving Latin American countries is relatively infrequent but when it does occur, it seems to primarily involve countries within the region.

\subsection*{Parameters with direct effect}

We first parameterize the model by identifying variables that we hypothesize have a direct impact on material conflict patterns between countries. There are a number of the standard explanations provided in the conflict literature. Inertia and reprocity top the list. Conflict in period $t$ is affected by what occurred previously in period $t-1$. This is autoregressive dependence. The expectation is that a dyad engaged in conflict in the previous period is more likely to be engaged in conflict in the next.

A lagged reciprocity parameter embodies the common argument that if country $j$ receives conflict from $i$ in period $t$, that in period $t+1$ $j$ may retaliate by sending conflict to $i$. The argument that reciprocity is likely to occur in conflict networks is certainly not novel, and has its roots in well known theories involving cooperation and conflict between states \citep{richardson:1960,choucri:north:1972,goldstein:1992}.

A number of exogenous explanations have often been used to explain conflicts between dyads. One of the most common relates to the role of geography. Apart from conflict involving major powers, conflict between countries that are geographically proximate is typical \citep{bremer:1992,diehl:goertz:2000,carter:goemans:2011}. Figure~\ref{fig:icews} demonstrates evidence of regional conflict patterns, as indicated by the clustering of similarly colored nodes, which represent countries within the same geographic region. This clustering suggests that conflicts are more likely to occur between countries in the same region. We use the minimum, logged distance between the dyads to operationalize this explanation.\footnote{Minimum distance estimation was conducted using the \pkg{CShapes} package \citep{weidmann:etal:2010a}.}

One of the most well developed arguments linking conflict between dyads to domestic institutions involves the idea of the democratic peace. The specific vein of this argument that has found the most support is the idea that democracies are unlikely to go to war with one another \citep{small:singer:1976,maoz:abdolali:1989,russett:oneal:2001}. Arguments for why democracies may have more peaceful relations between themselves range from how they share certain norms that make them less likely to engage in conflict to others hypothesizing that democratic leaders are better able to demonstrate resolve thus reducing conflict resulting from incomplete information \citep{maoz:russett:1993,fearon:1995}. To operationalize this argument, we construct a binary indicator that is one when both countries in the dyad are democratic.\footnote{We define a country as democratic if its polity score is greater than or equal to seven according to the Polity IV project \citep{marshall:jaggers:2002}.}

We also control for whether or not a pair of countries are allied to one another using data from the Correlates of War \citep{gibler:sarkees:2004}.\footnote{We consider a pair of countries allied to one another if they share a mutual defense treaty, neutrality pact, or entente.} Typically, one would expect that states allied to one another are less likely to engage in conflict. Another common control in the conflict literature is the level of trade between a pair of countries. We estimate trade flows between countries using the International Monetary Fund (IMF) Direction of Trade Statistics \citep{imf:2012}. Incorporating the level of trade between countries speaks to a long debate on the role that economic interdependencies may play in mitigating the risk of conflict between states \citep{barbieri:1996,gartzke:etal:2001}.\footnote{The extant literature has employed a variety of parameterizations to test this hypothesis. At times, a measure of trade dependence is calculated and at others just a simple measure of the trade flows between a pair of countries. We show results for the latter parameterization but results are consistent if we utilize a measure of trade dependence.}

The last set of measures we use to predict dyadic conflict are derived from another ICEWS quad variable. Verbal cooperation counts the occurrence of statements expressing a desire to cooperate from one country to another.\footnote{An example of a verbal cooperation event sent from Turkey to Portugal is the following: ``\textit{Portugal will support Turkey's efforts to become a full member of the European Community, Portuguese President Mario Soares said on Tuesday}.''}  We include a lagged and reciprocal version of this variable to our specification. This monthly level measure of cooperation between states provides us with a thermometer measure of the relations between states that is measured at a low level of temporal aggregation.

\subsection*{Parameters defining influence patterns}

We next add covariates to the model to explain the influence patterns observed in the network. The SIR model introduces the ability to explain these patterns using an underlying regression model, which is jointly estimated with the parameters modeling $y_{ij}$ through the iterative procedure described earlier. Using the SIR model we can answer the following types of questions:

\begin{itemize}
    \item Do the actions of one country at time $t-1$ influence the actions directed towards another country at time $t$ within the network, as reflected in the influence parameters $a_{ii'}$ and $b_{jj'}$?
	\item Which factors explain the network effects embedded in the influence parameters $a_{ii'}$ and $b_{jj'}$, determining the impact of one country's actions at time $t-1$ on the subsequent actions towards another country at time $t$?
\end{itemize}

The first covariate added to the influence specification, is simply a control for the distance between countries.\footnote{This is operationalized similarly as above using data from \pkg{CShapes}.} A negative effect for the distance parameter in the case of sender influence would indicate that countries are likely to send conflictual actions to the same countries that their neighbors are sending conflictual actions too. In the case of receiver influence, a negative effect would indicate that countries are likely to be targeted by the same set of countries that their neighbors are receiving conflictual interactions from.

An argument that has received continuing attention in the political science literature is the role that alliances play in either mitigating or exacerbating the level of conflict in the international system. Some have argued that in the case of a conflict, a country's allies will join in to honor their commitments thus increasing the risks for a multiparty interstate conflict \citep{snyder:1984,leeds:2003,vasquez:rundlett:2016}. We would find evidence for this argument if the ally parameter in the case of sender influence was positive, as that would indicate that countries are more likely to initiate or increase the level of conflict with countries that their allies are in conflict with.

The next covariate we consider is the volume of trade between countries. Trade relationships are often seen as a stabilizing factor in international relations, under the premise that economic interdependence reduces the likelihood of conflict by raising the costs of disruption \citep{keohane:nye:1977,oneal:russett:1999}. In the context of sender influence, a negative effect for the trade parameter would suggest that countries are less likely to initiate conflict with the same targets as their trading partners, supporting the idea that trade can act as a deterrent to conflict. Conversely, in the case of receiver influence, a positive effect would indicate that countries receiving conflict from others may also be the targets of those same countries' trading partners, potentially due to tensions arising from competitive trade dynamics.

The final covariate we examine is the level of verbal cooperation between countries, as indicated by diplomatic communications or public statements of support. Verbal cooperation can signal strong diplomatic ties or shared interests, potentially influencing patterns of conflict and cooperation in the network \citep{dorussen:ward:2008}. In the case of sender influence, a positive effect for the verbal cooperation parameter would imply that countries are more likely to align their conflictual actions with those of countries with whom they have a high degree of verbal cooperation, possibly as a show of solidarity or shared strategy. For receiver influence, a positive effect would suggest that countries facing conflict from one state may also find themselves targeted by that state's verbal allies, indicating a broader alignment in the international system.

Table~\ref{tab:modspec} summarizes each of the covariates used to estimate the social influence regression on the material conflict variable from ICEWS.

\begin{table}[ht]
\centering
	\begin{tabular}{ccll}
		\hline\hline
		\multirow{4}{*}{\Large{$Z_{ijt}$}} &
		\multirow{4}{*}{\resizebox{.1\textwidth}{!}{\begin{tikzpicture}

	\begin{scope}[xshift=1cm, yshift=1cm]
	\node{
		\begin{tikzpicture}[scale=.5]
			 \draw[thin, black,fill=green3] (0,0) grid (4,4) rectangle (0,0) ;
		\end{tikzpicture}
	};
	\end{scope}

	\begin{scope}[xshift=.5cm, yshift=.5cm]
	\node[](green){
		\begin{tikzpicture}[scale=.5]
			 \draw[thin, black,fill=green2] (0,0) grid (4,4) rectangle (0,0) ;
		\end{tikzpicture}
	} ;
	\end{scope}

	\begin{scope}
	\node{
		\begin{tikzpicture}[scale=.5]
			 \draw[thin, black,fill=green1] (0,0) grid (4,4) rectangle (0,0) ;
		\end{tikzpicture}
	};
	\end{scope}
	
\end{tikzpicture}}}
		&
		Material Conflict$_{ij,t-1}$ & Ally$_{ij,t-1}$ \\
		~ & ~ & Material Conflict$_{ji,t-1}$ & Log(Trade)$_{ij,t-1}$ \\
		~ & ~ & Distance$_{ij,t-1}$ & Verbal Cooperation$_{ij,t-1}$ \\
		~ & ~ & Joint Democracy$_{ij,t-1}$ & Verbal Cooperation$_{ji,t-1}$	\\
		\hline
		\multirow{4}{*}{\Large{$W_{ijt}$}} &
		\multirow{4}{*}{\resizebox{.1\textwidth}{!}{\begin{tikzpicture}

	\begin{scope}[xshift=1cm, yshift=1cm]
	\node{
		\begin{tikzpicture}[scale=.5]
			 \draw[thin, black,fill=red3] (0,0) grid (4,4) rectangle (0,0) ;
		\end{tikzpicture}
	};
	\end{scope}

	\begin{scope}[xshift=.5cm, yshift=.5cm]
	\node[](green){
		\begin{tikzpicture}[scale=.5]
			 \draw[thin, black,fill=red2] (0,0) grid (4,4) rectangle (0,0) ;
		\end{tikzpicture}
	} ;
	\end{scope}

	\begin{scope}
	\node{
		\begin{tikzpicture}[scale=.5]
			 \draw[thin, black,fill=red1] (0,0) grid (4,4) rectangle (0,0) ;
		\end{tikzpicture}
	};
	\end{scope}
	
\end{tikzpicture}}} & Distance$_{ij,t-1}$ \\
		~ & ~ &  Ally$_{ij,t-1}$ \\
		~ & ~ & Log(Trade)$_{ij,t-1}$  \\
		~ & ~ & Verbal Cooperation$_{ij,t-1}$ \\
		\hline\hline
	\end{tabular}
	\caption{Model specification summary for social influence regression. Top row shows covariates used to estimate direct effects and bottom sender and receiver influence.}
	\label{tab:modspec}
\end{table}
\FloatBarrier

\subsection*{Parameter Estimates}

\begin{figure}[ht]
	\centering
	\includegraphics[width=1\textwidth]{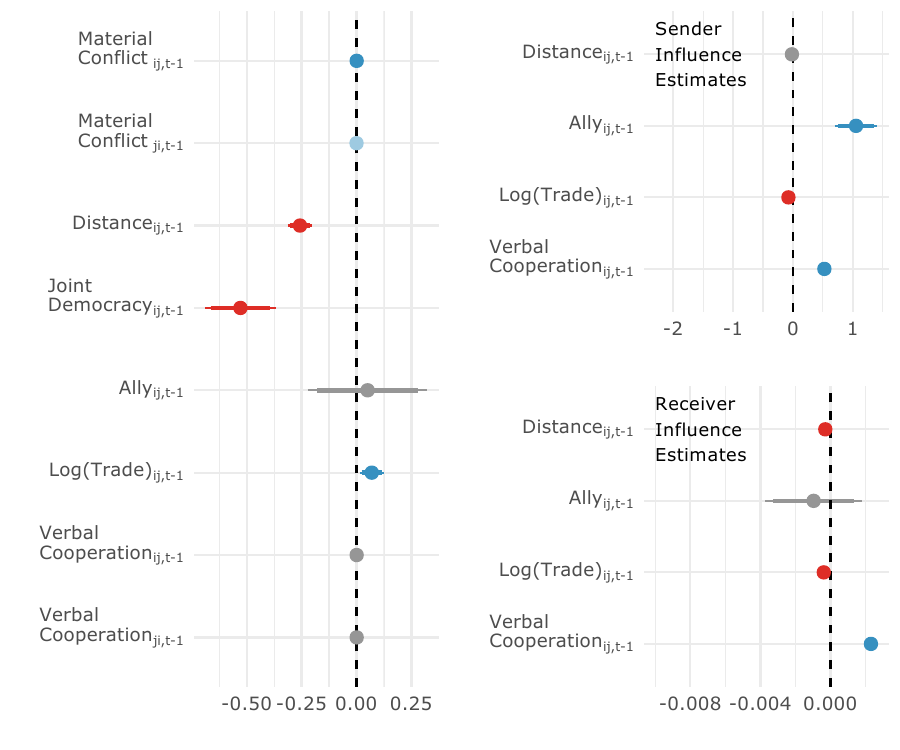}
	\caption{Left-most plot shows results for the direct effect parameters and the top-right plot represents results for the sender influence, and bottom-right receiver influence parameters. Points in each of the plots represents the average effect for the parameter and the width the 90 and 95\% confidence intervals. Dark shades of blue and red indicate that the parameter is significant at a 95\% confidence interval and lighter shades a 90\% confidence interval. Parameters that are not significant are shaded in grey.}
	\label{fig:correlConf}
\end{figure}

Figure~\ref{fig:correlConf} depicts the parameter estimates using a set of coefficient plots.\footnote{Convergence diagnostics are presented in Figure~\ref{fig:zabConv} of the Appendix.} On the left, we summarize the estimates of the direct effect parameters. As expected, greater levels of conflict between a dyad in the last period are associated with greater levels of conflict in the present. This speaks to a finding common in the conflict literature regarding the persistence of conflicts between dyads \citep{brandt:etal:2000}. We also find evidence that countries retaliate to conflict aggressively, though this effect is imprecisely measured. In terms of our exogenous parameters, the level of conflict between a dyad is negatively associated with the distance between them, a finding that aligns well with the extant literature.

Additionally, as is typical in the extant literature we find that jointly democratic dyads are unlikely to engage in conflict with one another. Surprisingly, however,  the level of trade between countries is positively associated with the level of conflict. The divergence of this finding with some of the extant literature may be a result of a variety of factors, such as our use of a measure of conflict that has much greater variance than the militarized interstate disputes measurement from the Correlates of War dataset. 

\FloatBarrier

The right-most plots focuses on sender (top) and receiver (bottom) influence patterns. Notably, the alliance sender influence parameter has a positive effect, indicating that countries tend to initiate greater levels of conflict with countries that their allies were fighting in the previous period. This finding is in line with arguments in the extant literature about the role that alliance relationships may play in leading to more conflict in the international system \citep{siverson:king:1980,leeds:2005}.

Additionally, countries are likely to send conflict to those with whom their verbal cooperation partners are initiating or increasing conflict with. This finding is interesting as it highlights that countries making cooperative statements regarding a particular country $i$, actually go beyond those statements in later periods to supporting $i$ by initiating conflict with those that $i$ was in conflict with. Trade flows, on the other hand, are associated with having a negative effect, implying that countries are not likely, and in fact somewhat unlikely, to follow their trading partners into conflict.

Receiver influence patterns are similarly determined. Trade flows and verbal cooperation have similar effects, though the interpretation here for trade is that countries are unlikely to be targeted by those that target their trading partners. Interestingly, the distance effect on the receiver influence side is more precisely measured, implying that geographically proximate countries are more likely to receive conflict from a similar set of countries.

\subsection*{Visualizing Dependence Patterns}

\begin{figure}[ht]
	\centering
	\begin{tabular}{l}
		\scshape{\textbf{June 2007}} \\
		\scshape{\;\;\;\; Sender Dependence Patterns} \\
		\includegraphics[width=.8\textwidth]{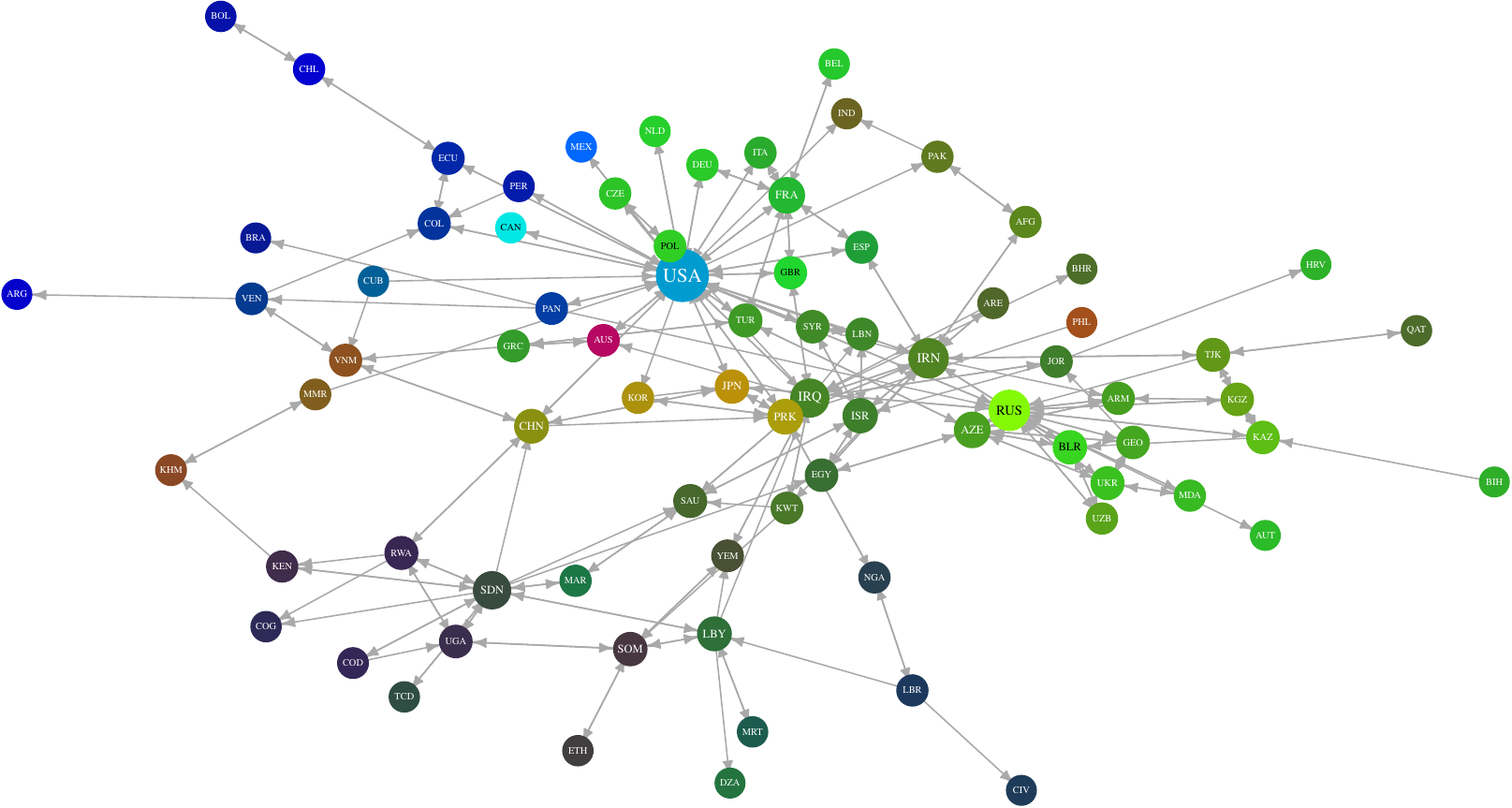} \\
		\scshape{\;\;\;\; Receiver Dependence Patterns} \\
		\includegraphics[width=.8\textwidth]{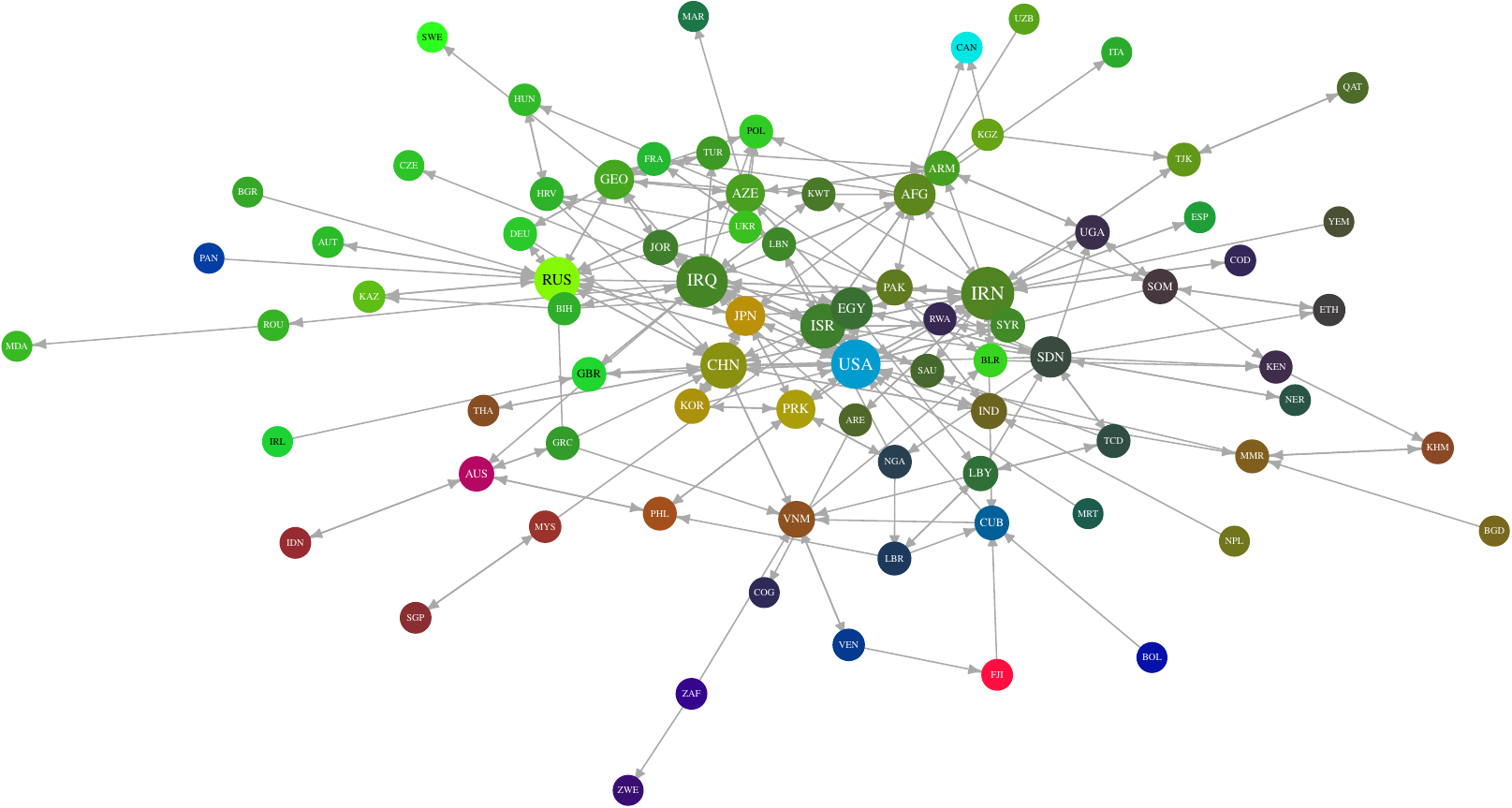} \\
	\end{tabular}
	\caption{Network visualization of influence patterns as estimated by the social influence regression model for June 2007. Nodes are colored by their relative geographic position and are sized by the number of influence relationships that they receive and send.}
	\label{fig:inflRel}
\end{figure}

Based on the sender and receiver influence parameter estimates, Figure~\ref{fig:inflRel} provides a visual summary of the type of dependence patterns that are implied in the context of the material conflict model estimated in the previous section.

The linear combination of our influence parameter estimates ($\alpha$), and the design array containing sender influence variables ($w_{ijt}$) are used to visualize the sender dependence patterns between a pair of countries ($a_{ijt}$): $a_{ijt} = \alpha^{\top} w_{ijt}$. The resulting sender and receiver dependence pattern are shown in Figure~\ref{fig:inflRel} for June 2007.\footnote{A lengthier table of visualizations for additional time periods is shown in Figure~\ref{fig:inflRelLong} of the Appendix.} For the visualization on the left [right], edges between countries indicate that greater likelihood to send [receive] conflictual events to [from] the same countries. Countries are colored by their relative geographic position and node size corresponds to the number of influence relationships the country shares.

Since these dependence patterns are estimated directly from the model results that are presented in Figure~\ref{fig:correlConf}, the patterns implied by that model are manifest in these visualizations. One of the more notable findings from the sender influence model is the role that alliance relationships play, and this effect is striking. For example, the USA shares sender influence ties with a number of Western European countries such as Germany and the United Kingdom, the USA also is more likely to send conflict to actors that Australia, South Korea and Japan have engaged in material conflict with, and many of these countries are likely to do the same.

\vspace{-2mm}
\FloatBarrier

A predictor of receiver influence patterns is the distance between countries. Countries are more likely to be targeted by the same set of countries as their neighbors. This pattern manifests itself in the right-most visualization in Figure~\ref{fig:inflRel}, where we find clumps of countries, such as Iraq, Lebanon, and Jordan, clustering together.

\subsection*{Performance Comparison}

A common and important argument for employing a network based approach is that it aids in better accounting for the data generating process underlying relational data structures. Thus, in this case, the network approach should actually better predict conflict in an out-ofsample test. To put the performance of this model in context, we compare it to a standard GLM that does not account for dependence patterns in the network, but is similarly parameterized. Additionally, given the recent interest in machine learning methods as tools for prediction within the social sciences we compare the performance against a generalized boosted model (GBM).

Boosting methods have become a popular approach in the machine learning to ensemble over decision tree models in a sequential manner. At each iteration, a new model is trained with respect to the error of the ensemble at that point. \citet{friedman:2001} greatly extended the learning procedure underlying boosting algorithms, by modifying the approach to choose new models at every iteration so that they would be maximally correlated with the negative gradient of some loss function relevant to the ensemble. In the case of a squared-error loss function, this would correspond to sequentially fitting the residuals. We use a generalized version of this model developed by \citet{ridgeway:2012} that extends this framework to the estimation of a variety of distribution types---in our case, a Poisson regression model. In general, these types of models have been shown to give substantial predictive advantage over alternative methods, such as GLM, and should provide a useful point of comparison.\footnote{The $\sf{R}$ \pkg{gbm} package on CRAN implements this estimator \citep{ridgeway:2012}.}

To compare these approaches we first utilize a cross-validation procedure. This involves first randomly dividing $T$ time points in our relational array into $k=10$ sets and within each set we set randomly exclude five time slices from our material conflict array. We then run our models and predict the five missing slices from the estimated parameters. Proper scoring rules are used to compare predictions. Scoring rules evaluate forecasts through the assignment of a numerical score based on the predictive distribution and on the actual value of the dependent variable. \citet{czado:etal:2009} discuss a number of such rules that can be used for count data: Brier, Dawid-Sebastiani, Logarithmic, and Spherical scores.\footnote{Details are provided in the Appendix.} For each of these rules, lower values on the metric indicate better performance.

\begin{figure}[ht]
	\centering
	\includegraphics[width=.8\textwidth]{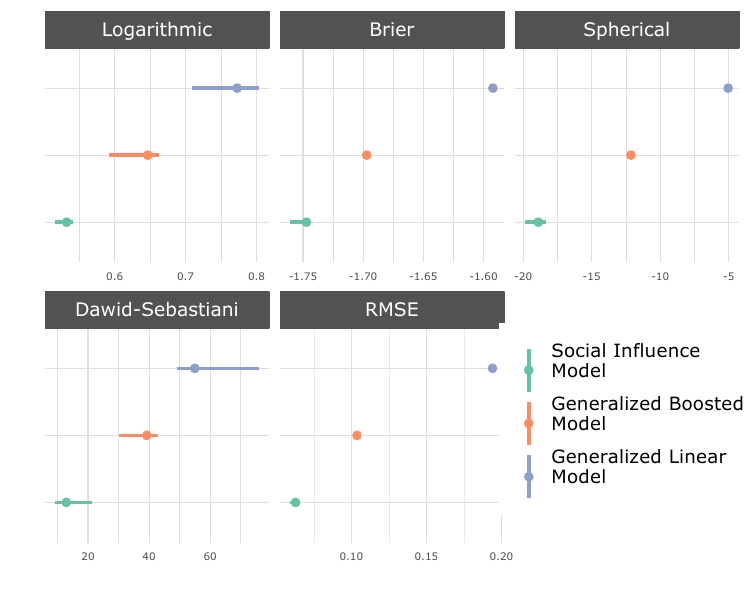}
	\caption{Performance comparison based on randomly excluding time slices from the material conflict array. Colors designate the different models, and the average score across the 10-fold cross validation is designated by a circle and the range by a horizontal line.}
	\label{fig:predCompareRandom}
	\end{figure}
\FloatBarrier

Figure~\ref{fig:predCompareRandom} illustrates differences in the performance between the social influence model, GLM, and GBM across the scoring rules mentioned above and a more standard metric, the RMSE. In the case of each of these metrics we find GLM performs the worst and that the social influence model performs the best.

\begin{figure}[ht]
	\centering
	\includegraphics[width=.8\textwidth]{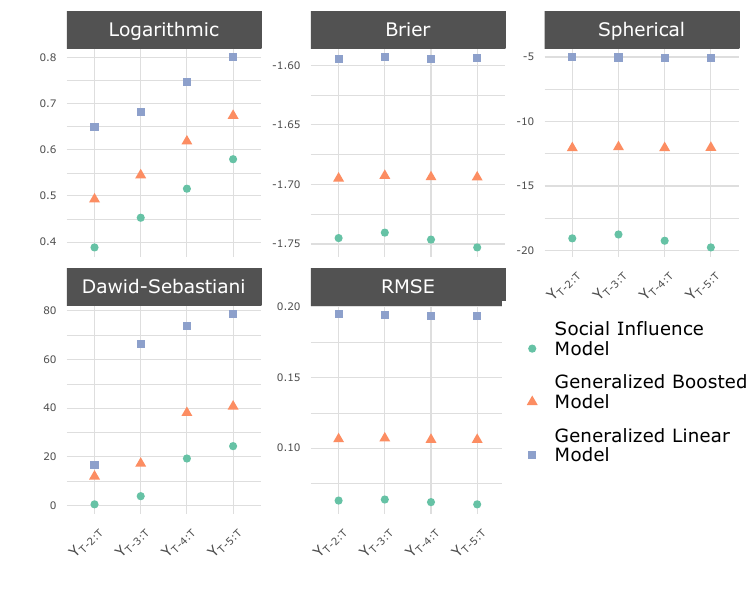}
	\caption{Performance comparison based on randomly excluding the last two to five periods of the material conflict array. Colors and shapes designate the different models, and the score when excluding $x$ number of periods is shown.}
	\label{fig:predCompareTime}
	\end{figure}

We also assess the predictive accuracy of our models in a forecasting context. We perform such an exercise as well by dividing up our sample into a training and test set, where the test set corresponds to the last $x$ periods in the data that we have available. We vary $x$ from two to five. For instance, when $x=5$ we are leaving the last five years of data for validation. Results for this analysis are shown in Figure~\ref{fig:predCompareTime} and there again we find that the social influence model has better out of sample predictive performance than the alternatives we test here.

\FloatBarrier

\section*{\textbf{Conclusion}}

In this paper, we introduced the Social Influence Regression (SIR) model, which represents an important extension of the bilinear network autoregression framework, designed to more effectively capture and explain influence dynamics within networks. The SIR model addresses a key limitation of existing models by incorporating exogenous covariates into the estimation process, allowing us to directly model and interpret the factors driving influence within a network. This approach not only enhances the explanatory power of the model but also provides a more rigorous and theoretically grounded framework for understanding complex relational data. A key contribution of our work is the development of a more efficient estimation scheme for the SIR model. Using an iterative block coordinate descent method, we enhance the model's computational feasibility, especially for large-scale networks. 

The application of the SIR model to the study of material conflict between countries provided several important insights that underscore the model's practical utility. By incorporating covariates such as geographic proximity, alliances, trade, and verbal cooperation, the SIR model revealed nuanced patterns of influence within the international conflict network. For example, the model identified that countries tend to initiate conflicts against the same targets as their allies, a finding that aligns with established theories in international relations about the role of alliances in escalating conflicts. Additionally, the negative influence of trade flows on conflict initiation suggested that countries are less likely to follow their trading partners into conflict, highlighting the stabilizing effect of economic interdependence. Verbal cooperation was shown to have a reinforcing effect, where countries that publicly support each other are more likely to align their conflictual actions. These findings not only validate the robustness of the SIR model but also demonstrate its ability to generate new theoretical and empirical insights into the dynamics of international conflict, providing a clearer understanding of the factors driving influence within complex networks.

Looking forward, the SIR model opens up numerous avenues for future research, both in terms of its applications and methodological developments. The model's flexibility allows it to be adapted to various network contexts beyond international conflict. Methodologically, there are several promising directions for refinement and extension. One area for development is the further optimization of the block coordinate descent method, particularly for handling even larger and more complex networks. This could involve parallelizing the estimation process or incorporating advanced optimization techniques such as stochastic gradient descent to improve scalability and convergence speed.

\clearpage

\renewcommand{\thefigure}{A\arabic{figure}}
\setcounter{figure}{0}
\renewcommand{\thetable}{A.\arabic{table}}
\setcounter{table}{0}
\renewcommand{\thesection}{A.\arabic{section}}
\setcounter{section}{0}

\section*{\textbf{Appendix}}

Visualization of convergence for direct (blue), sender influence (green), and receiver influence (red) parameters.

\begin{figure}[ht]
\centering
\includegraphics[width=1\textwidth]{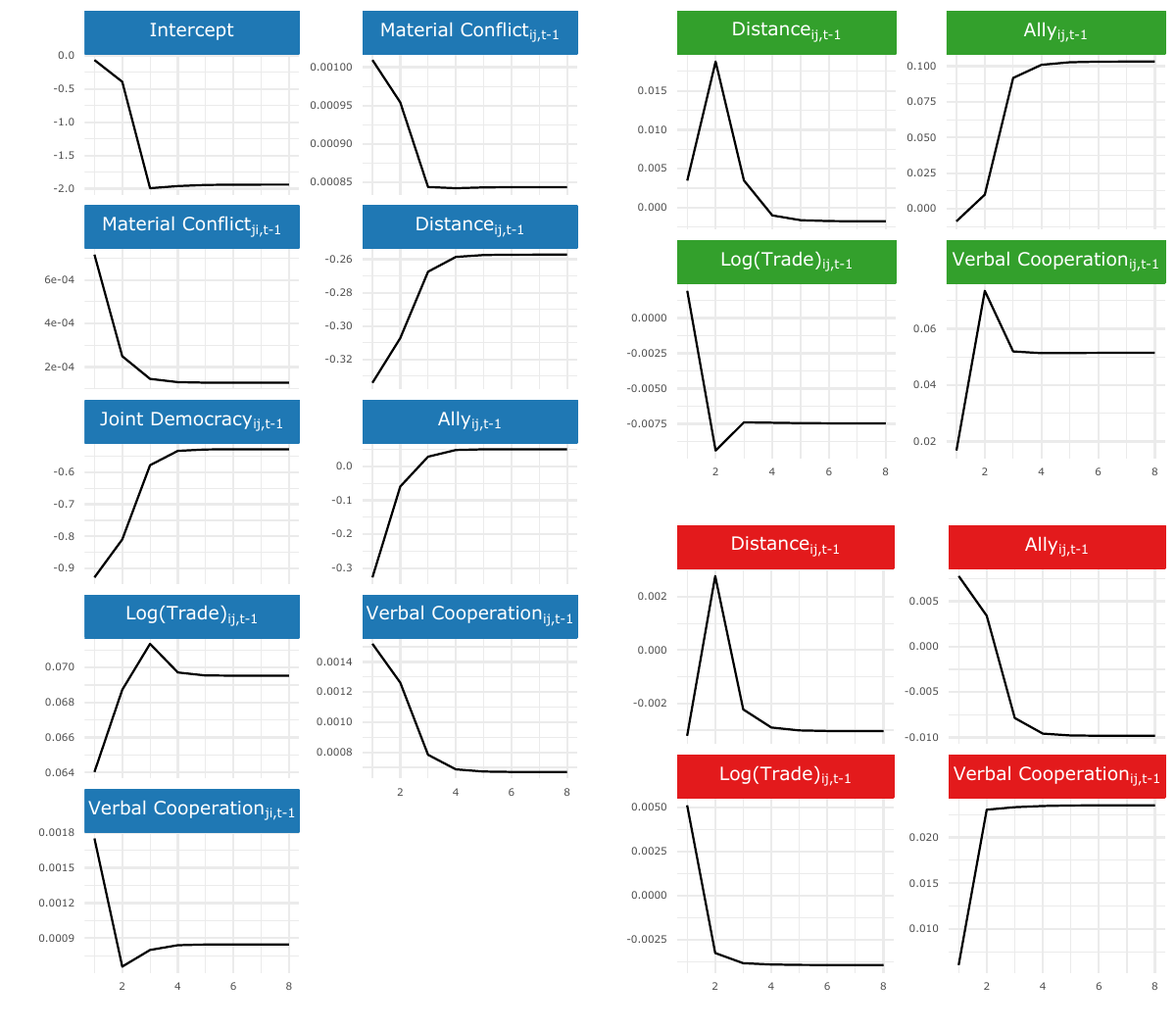}
\caption{Convergence diagnostics for the social influence regression model on material conflict.}
\label{fig:zabConv}
\end{figure}

\clearpage
\newpage
\subsection*{Convergence}

\subsection*{Influence Dynamics}

Visualization of influence effects for select time points from dynamic social influence regression model.

\begin{figure}[ht]
\centering
	\begin{tabular}{lcr}
		\scshape{\scriptsize{Sender Influence Space:}} & ~ & ~  \\
		\scshape{\tiny{February 2005}} & \scshape{\tiny{September 2006}} & \scshape{\tiny{June 2007}} \\
			\includegraphics[height=.1\textheight]{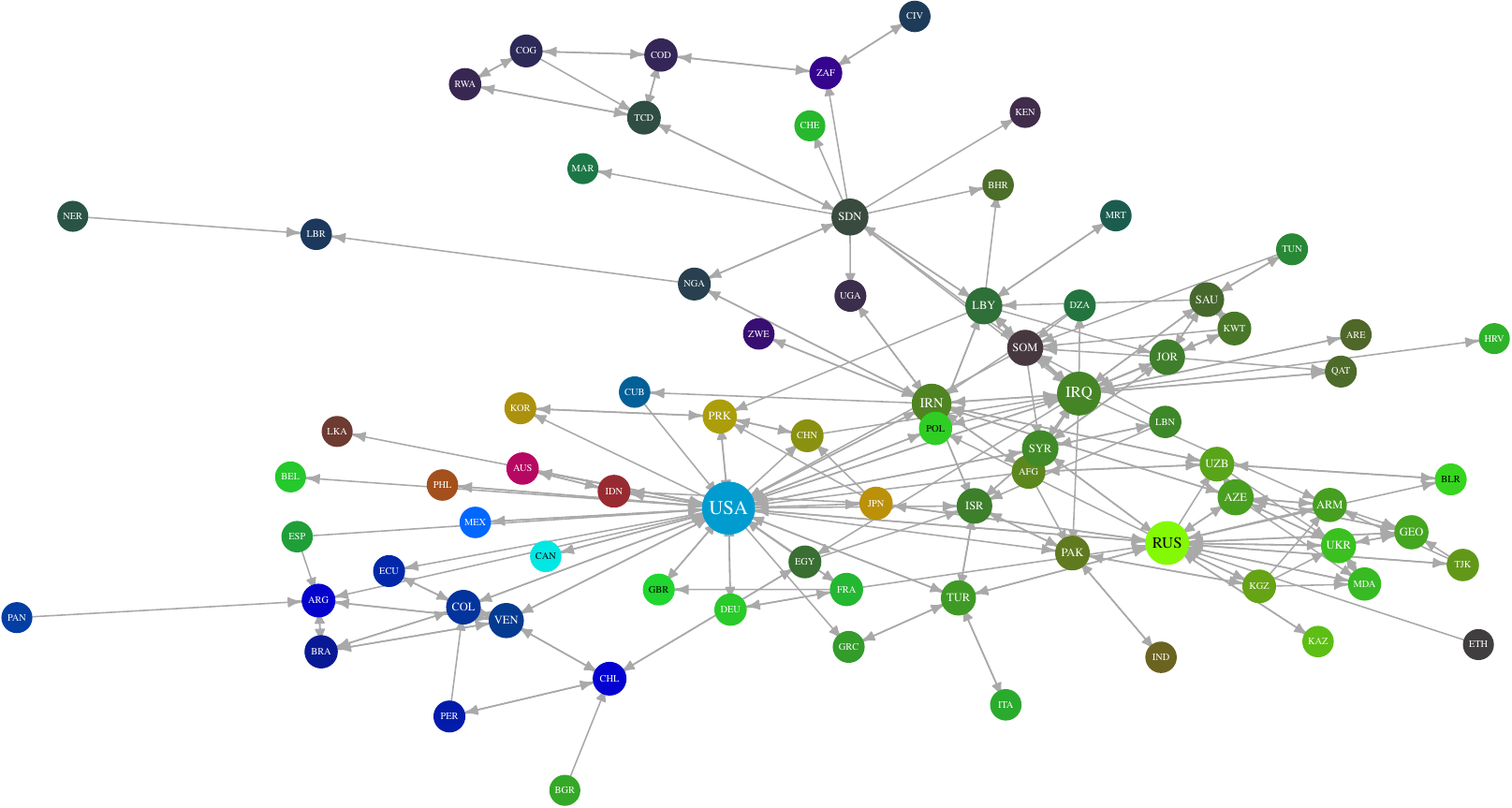} & 
			\includegraphics[height=.1\textheight]{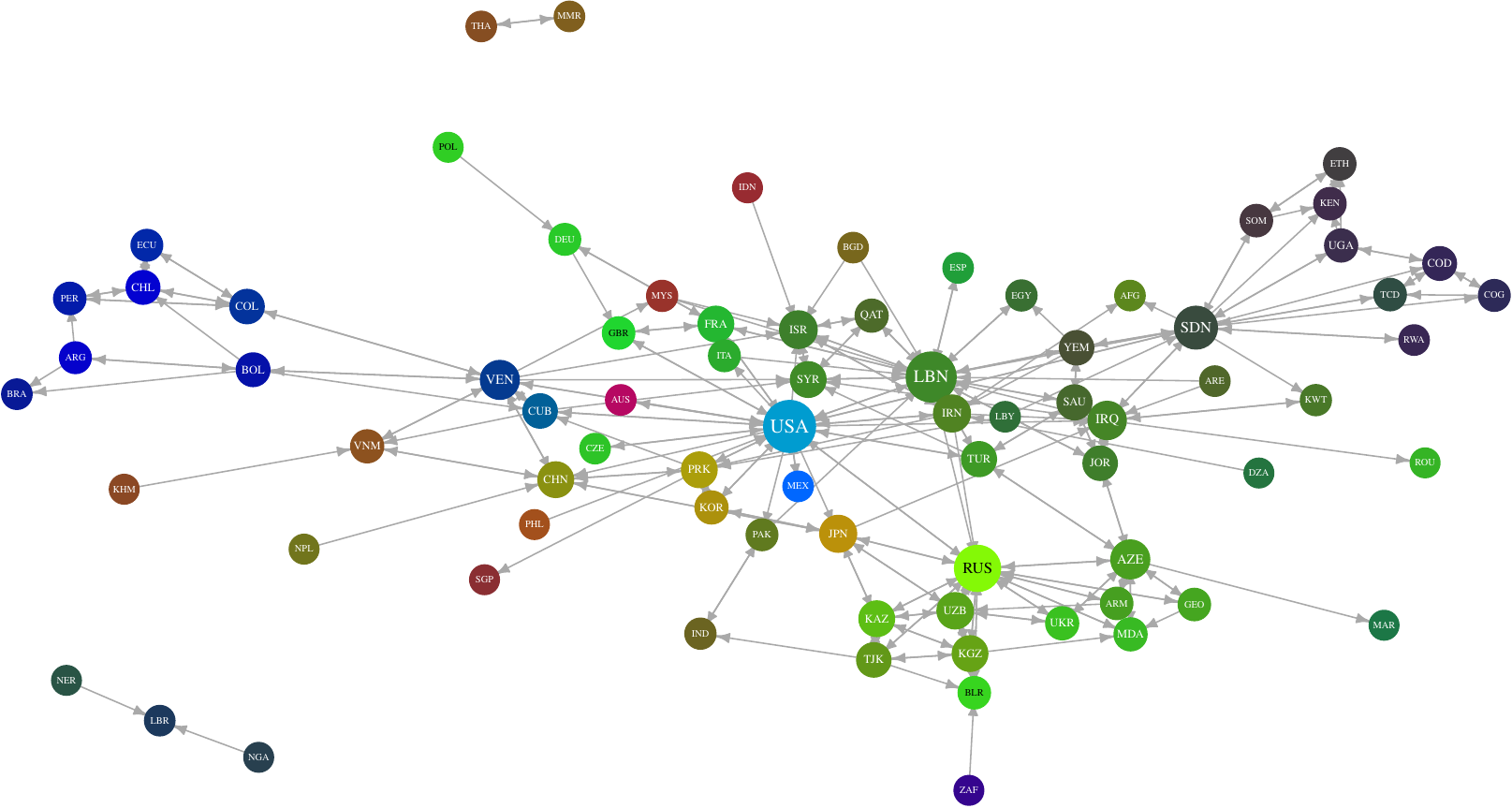} & 
			\includegraphics[height=.1\textheight]{aInfl_2007_06_01.pdf} \\
		\scshape{\tiny{April 2008}} & \scshape{\tiny{January 2009}} & \scshape{\tiny{August 2010}} \\
			\includegraphics[height=.1\textheight]{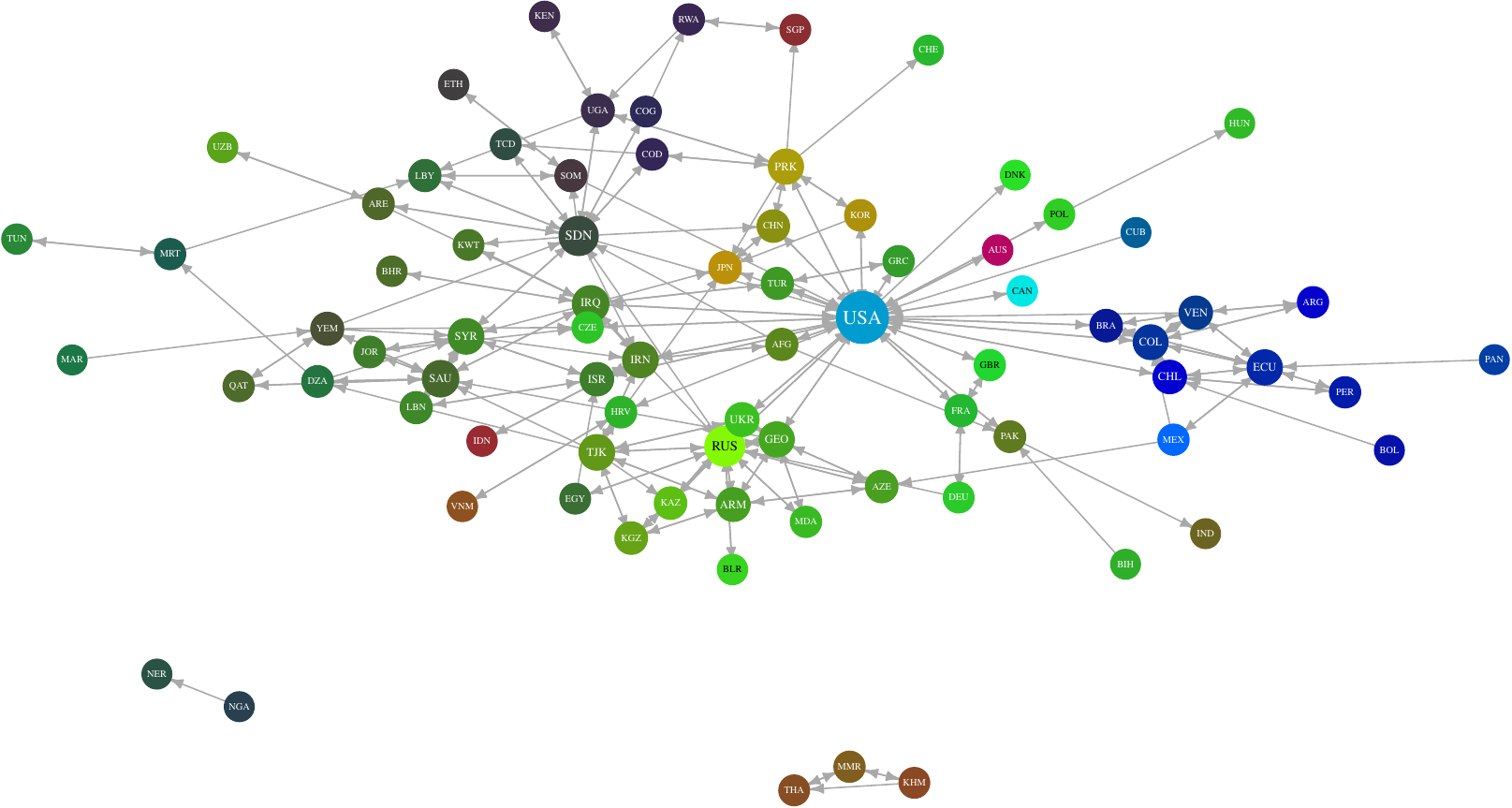} & 
			\includegraphics[height=.1\textheight]{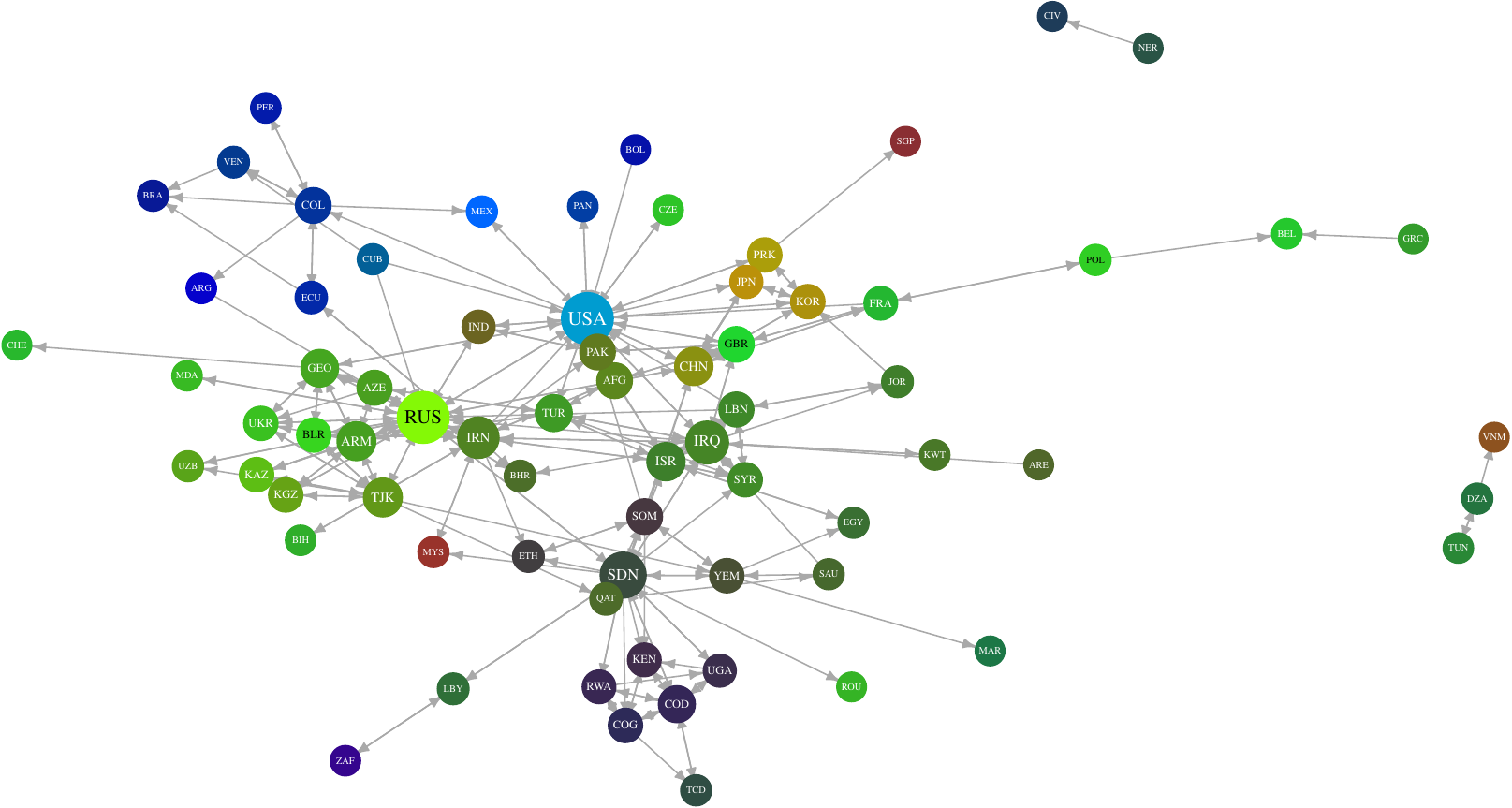} &
			\includegraphics[height=.1\textheight]{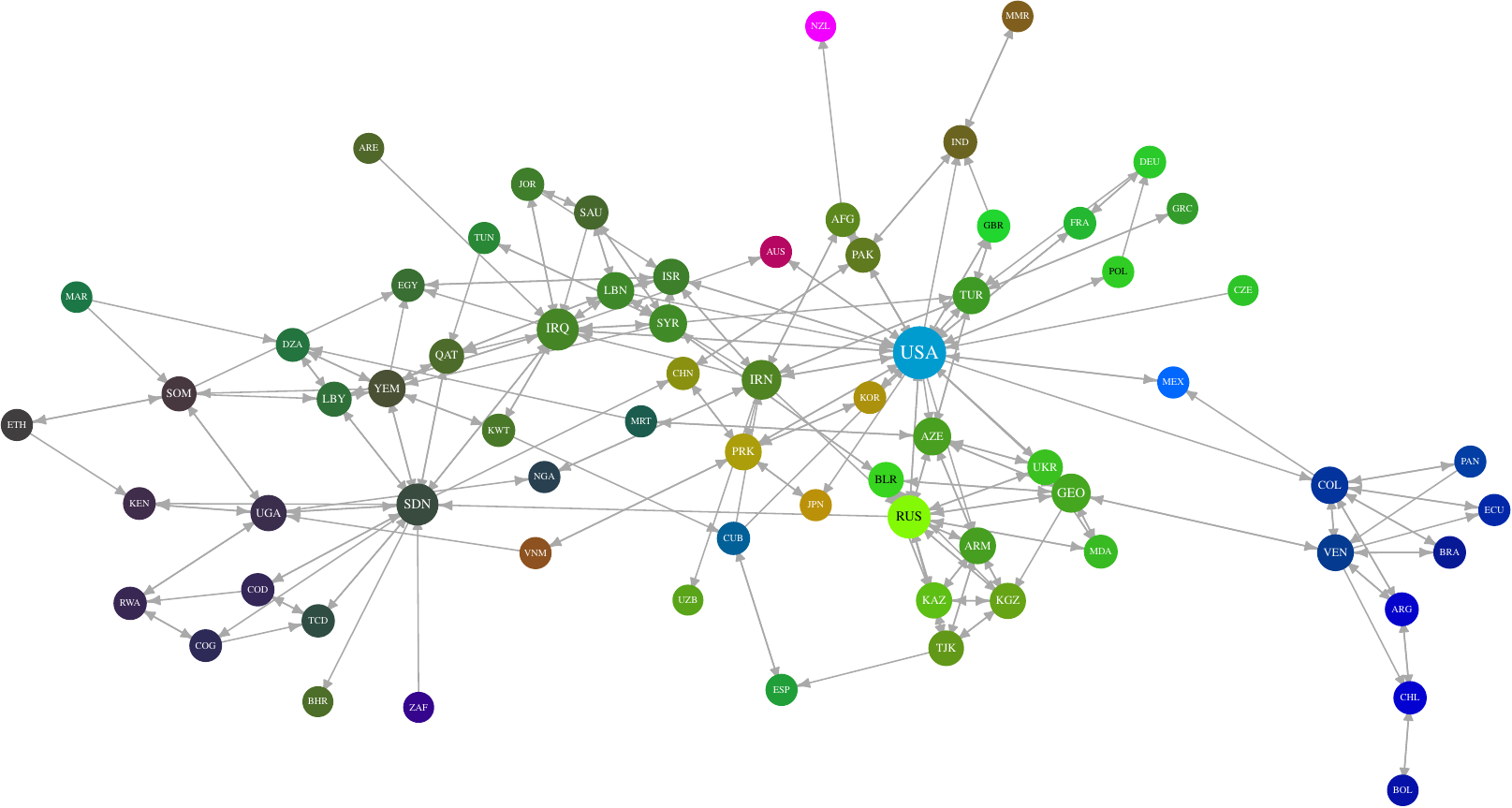} \\
		\scshape{\tiny{October 2009}}  & \scshape{\tiny{May 2011}} & \scshape{\tiny{December 2012}}\\
			\includegraphics[height=.1\textheight]{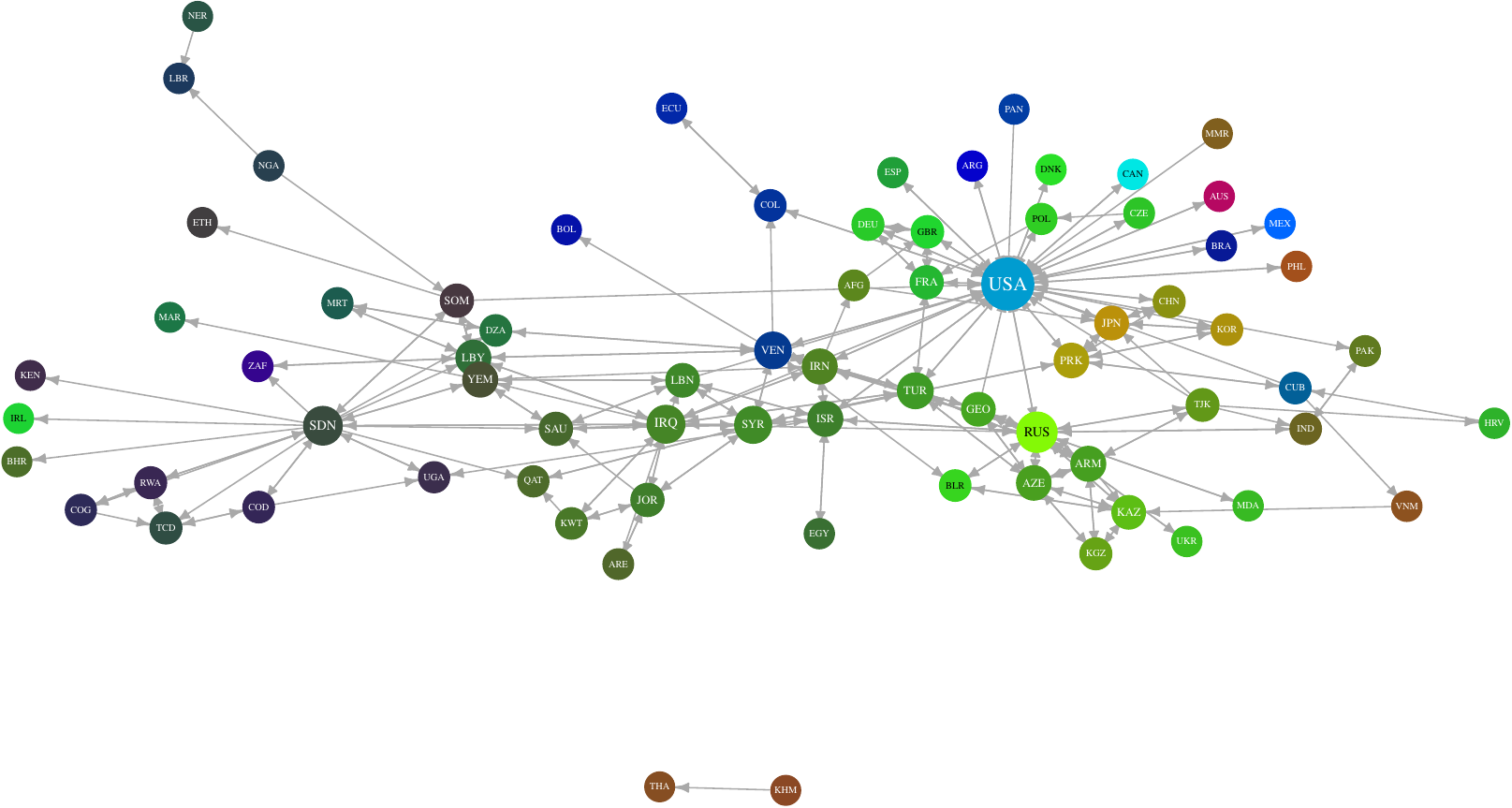} & 
			\includegraphics[height=.1\textheight]{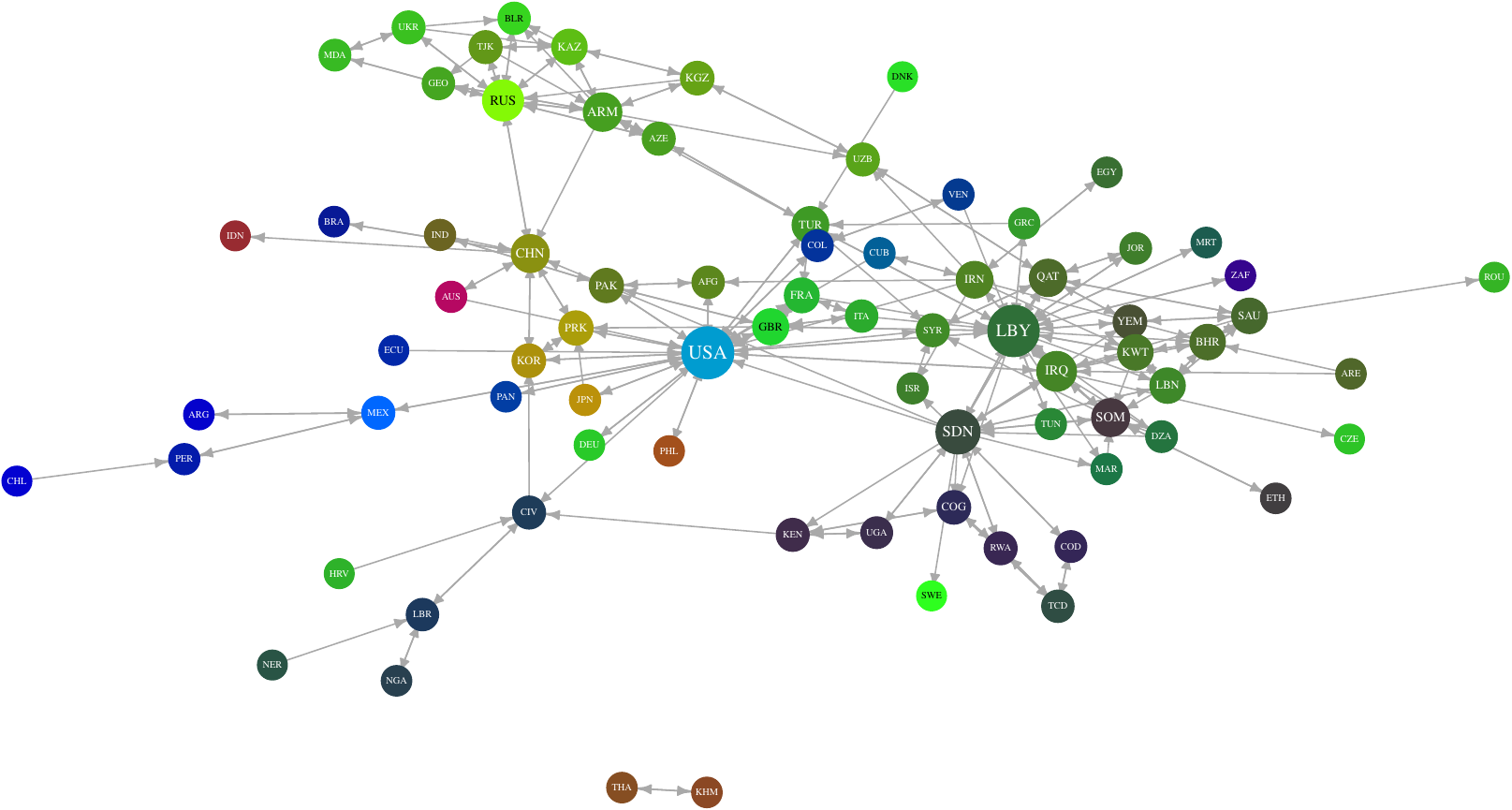} &		
			\includegraphics[height=.1\textheight]{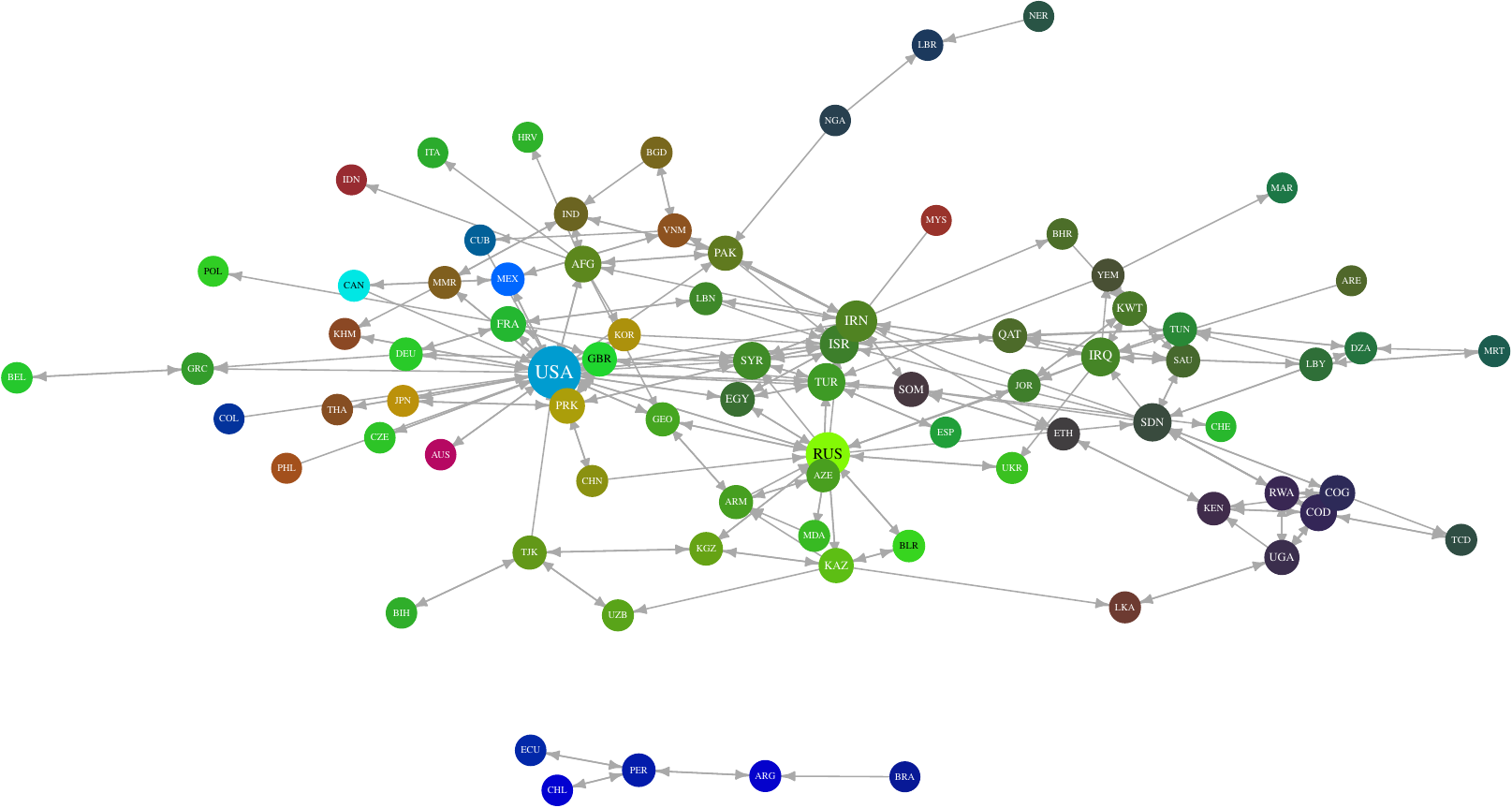} \\
		\scshape{\scriptsize{Receiver Influence Space:}} & ~ & ~  \\
		\scshape{\tiny{February 2005}} & \scshape{\tiny{September 2006}} & \scshape{\tiny{June 2007}} \\
			\includegraphics[height=.1\textheight]{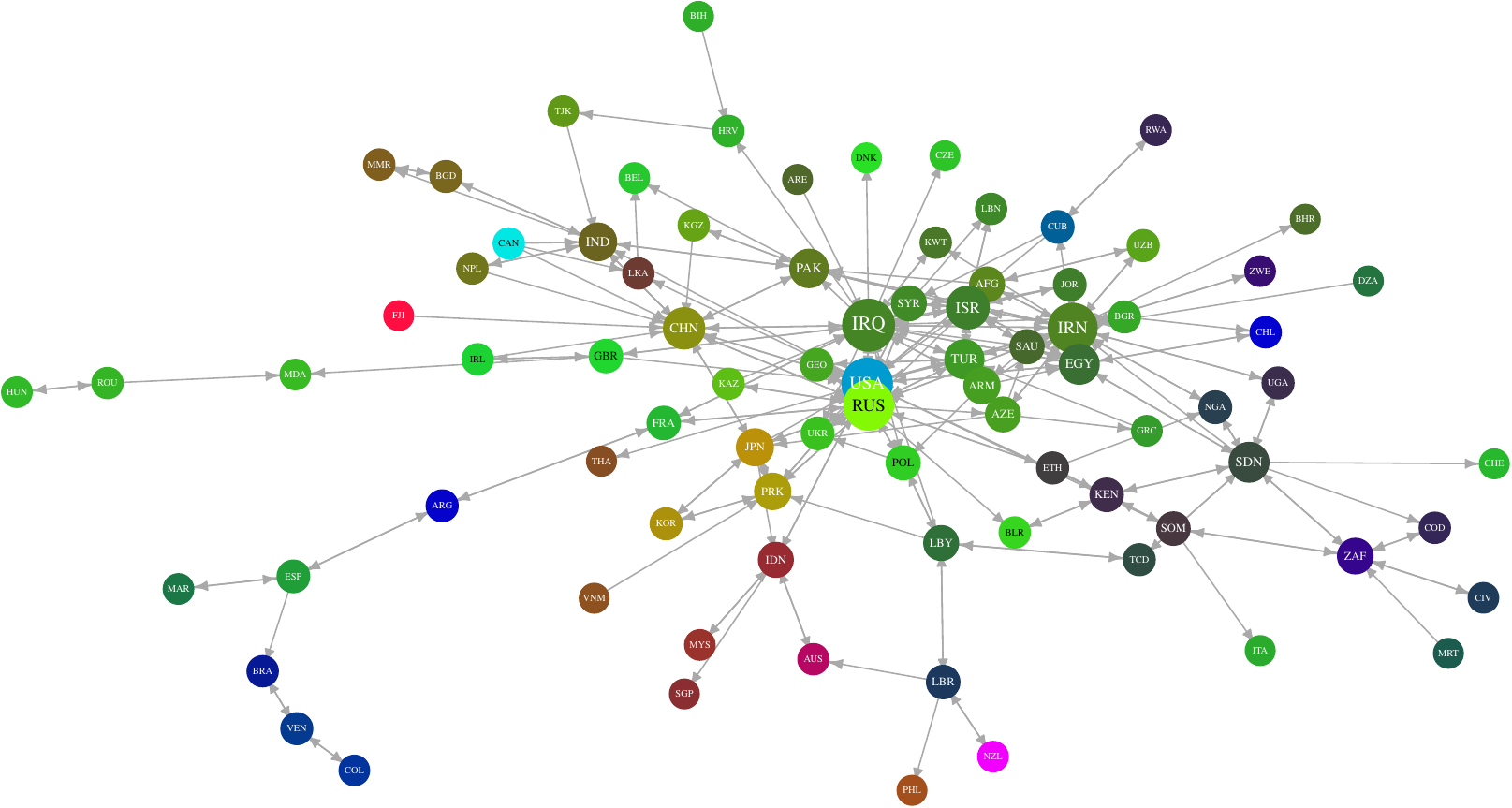} & 
			\includegraphics[height=.1\textheight]{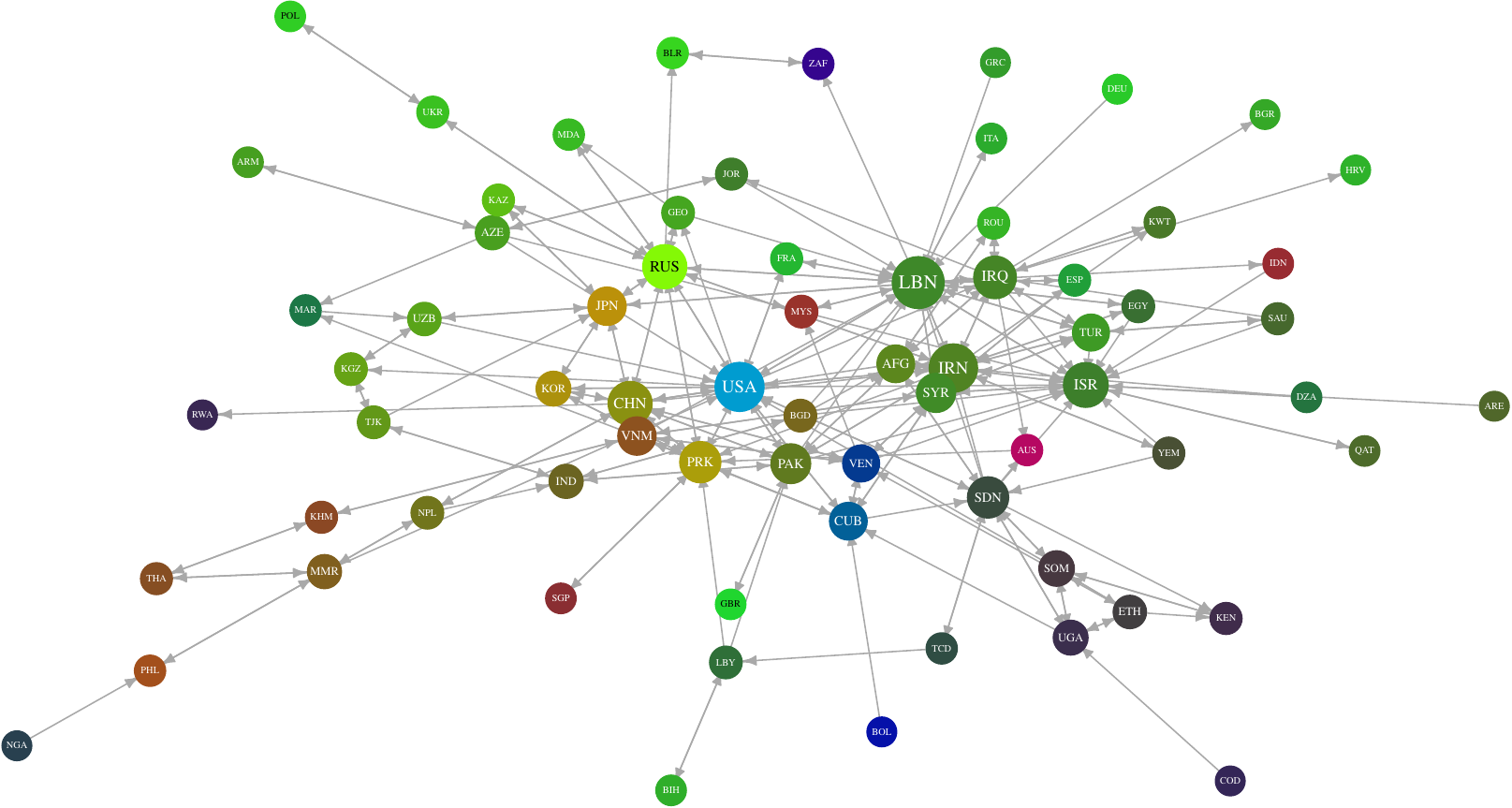} & 
			\includegraphics[height=.1\textheight]{bInfl_2007_06_01.pdf} \\
		\scshape{\tiny{April 2008}} & \scshape{\tiny{January 2009}} & \scshape{\tiny{August 2010}} \\
			\includegraphics[height=.1\textheight]{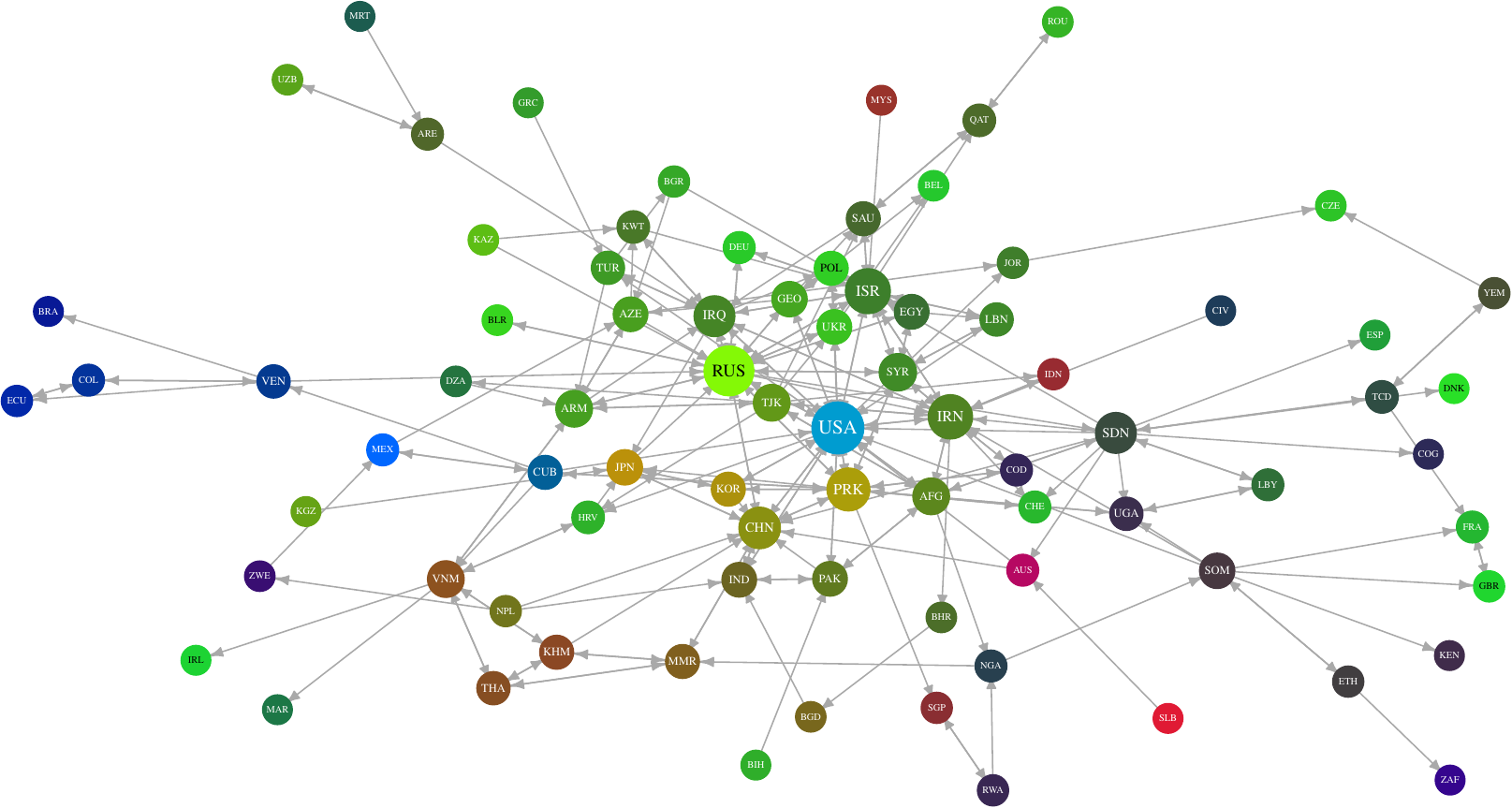} & 
			\includegraphics[height=.1\textheight]{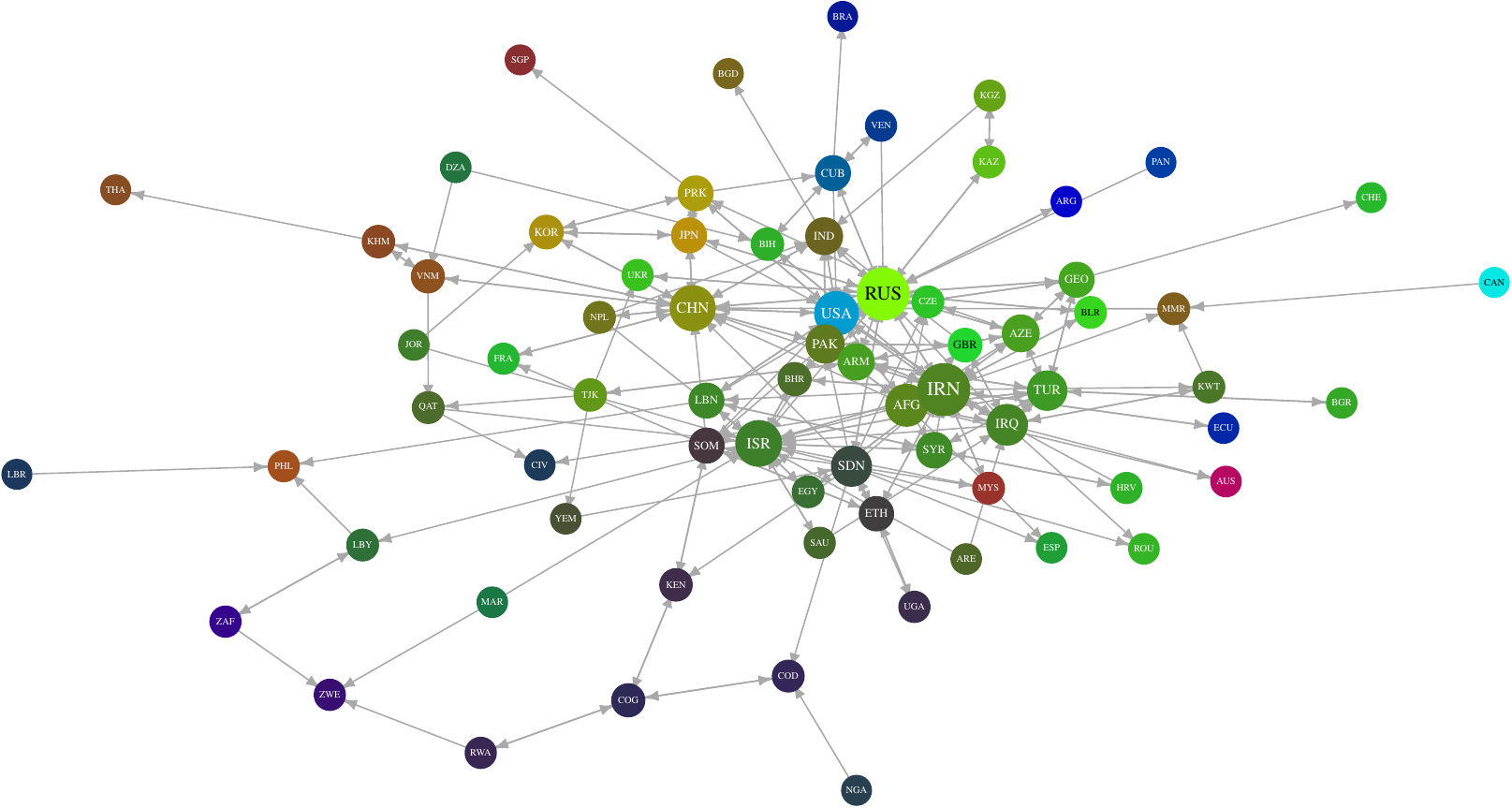} &
			\includegraphics[height=.1\textheight]{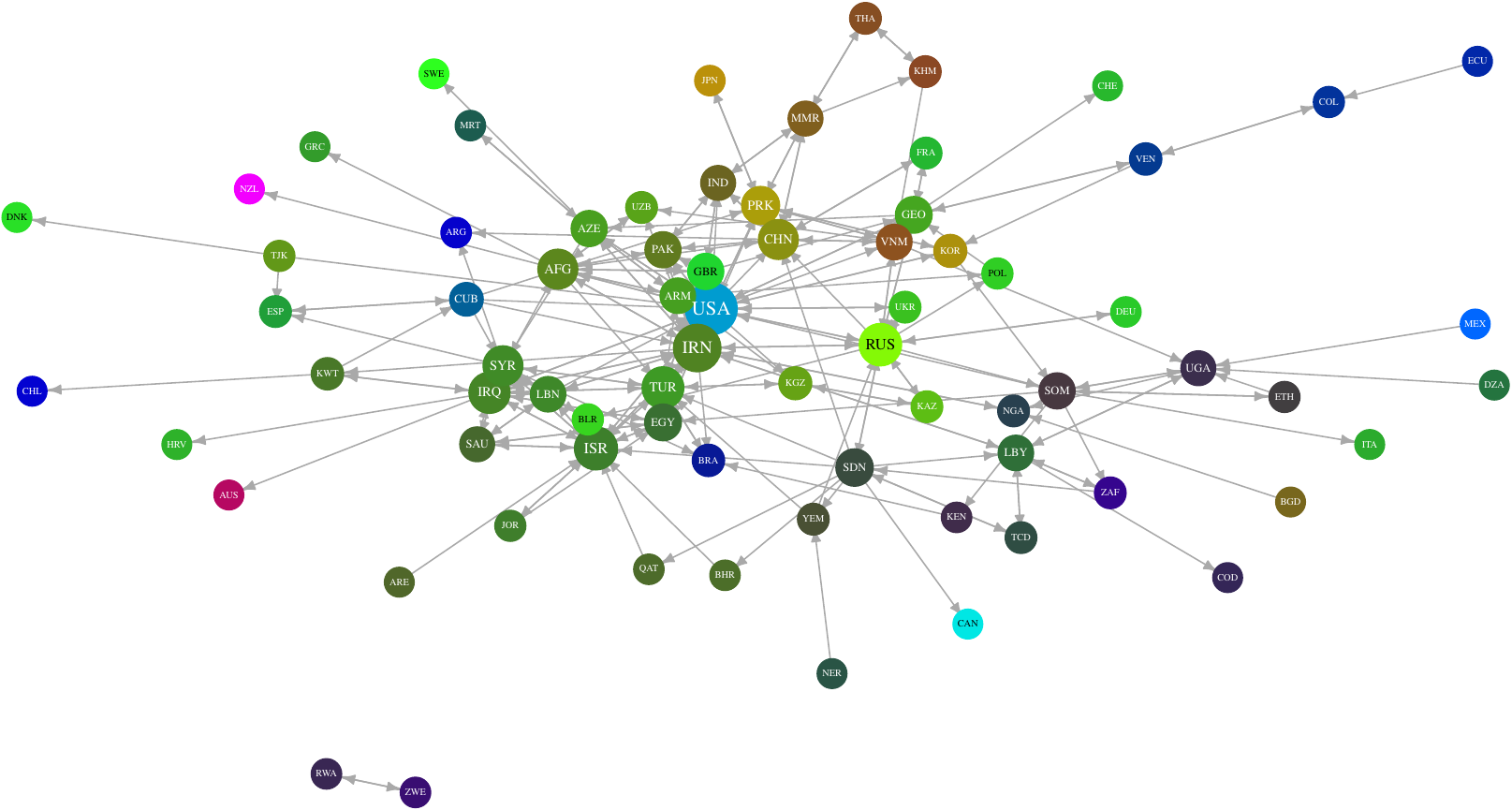} \\
		\scshape{\tiny{October 2009}}  & \scshape{\tiny{May 2011}} & \scshape{\tiny{December 2012}}\\
			\includegraphics[height=.1\textheight]{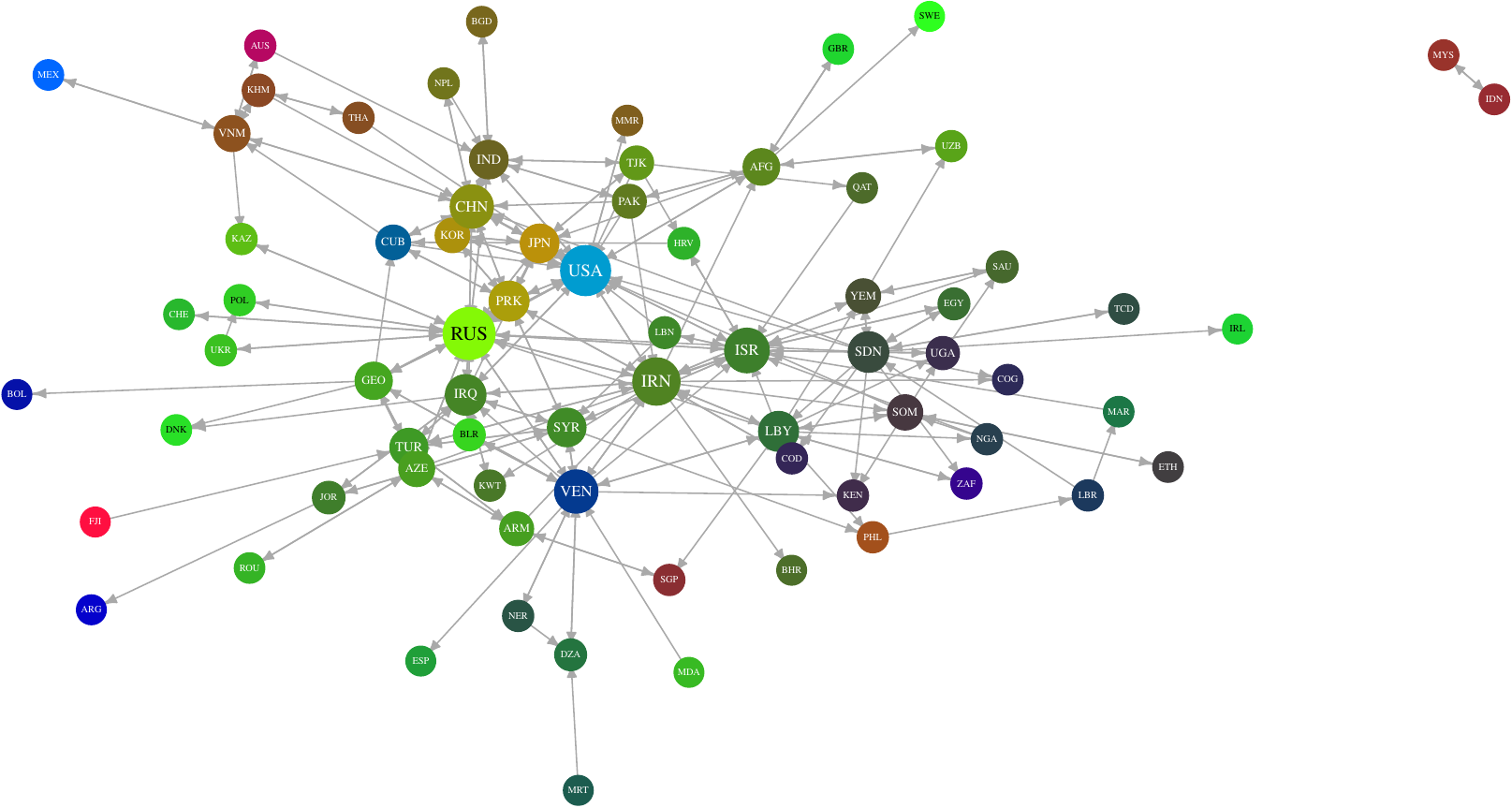} & 
			\includegraphics[height=.1\textheight]{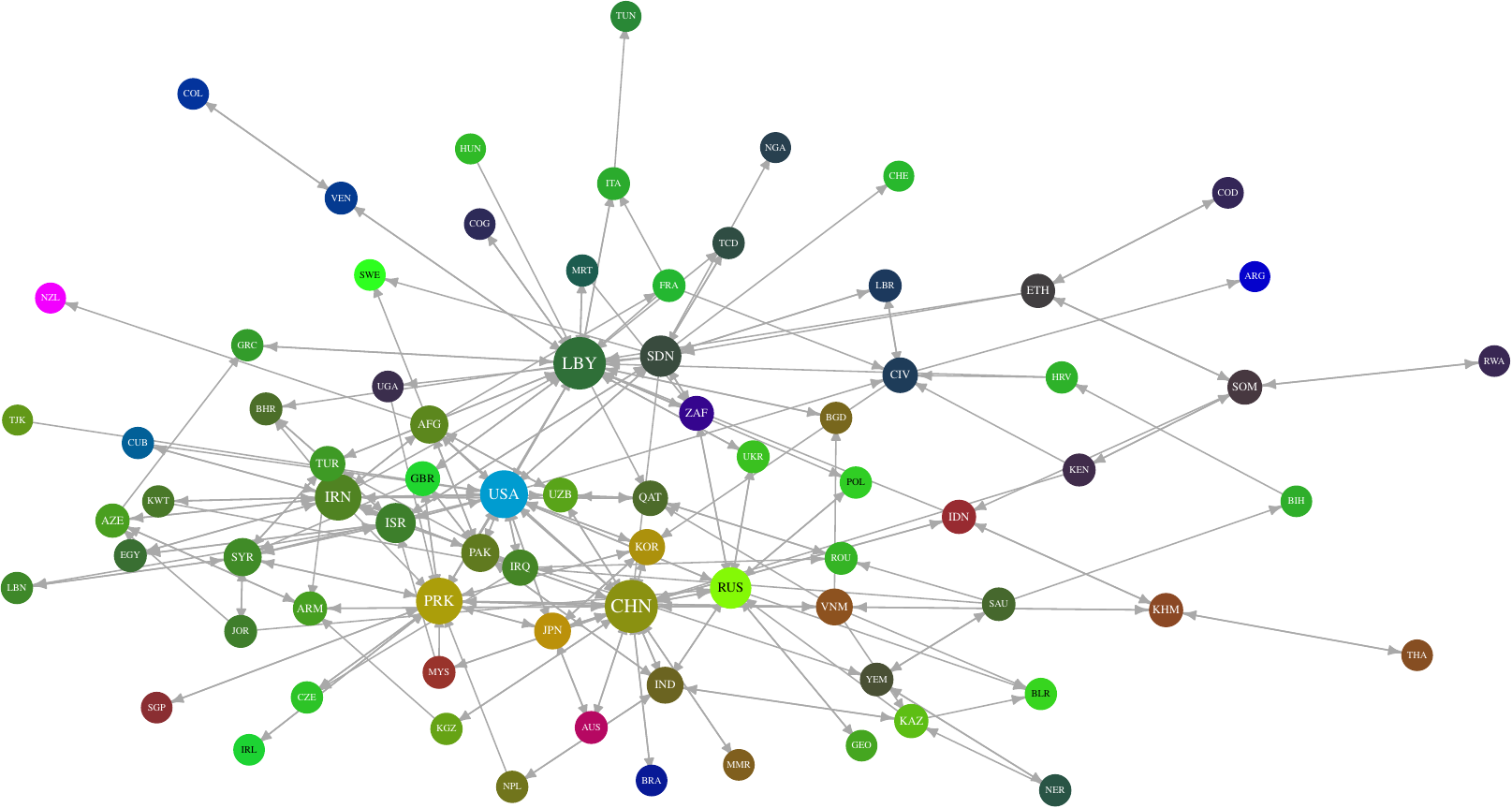} &		
			\includegraphics[height=.1\textheight]{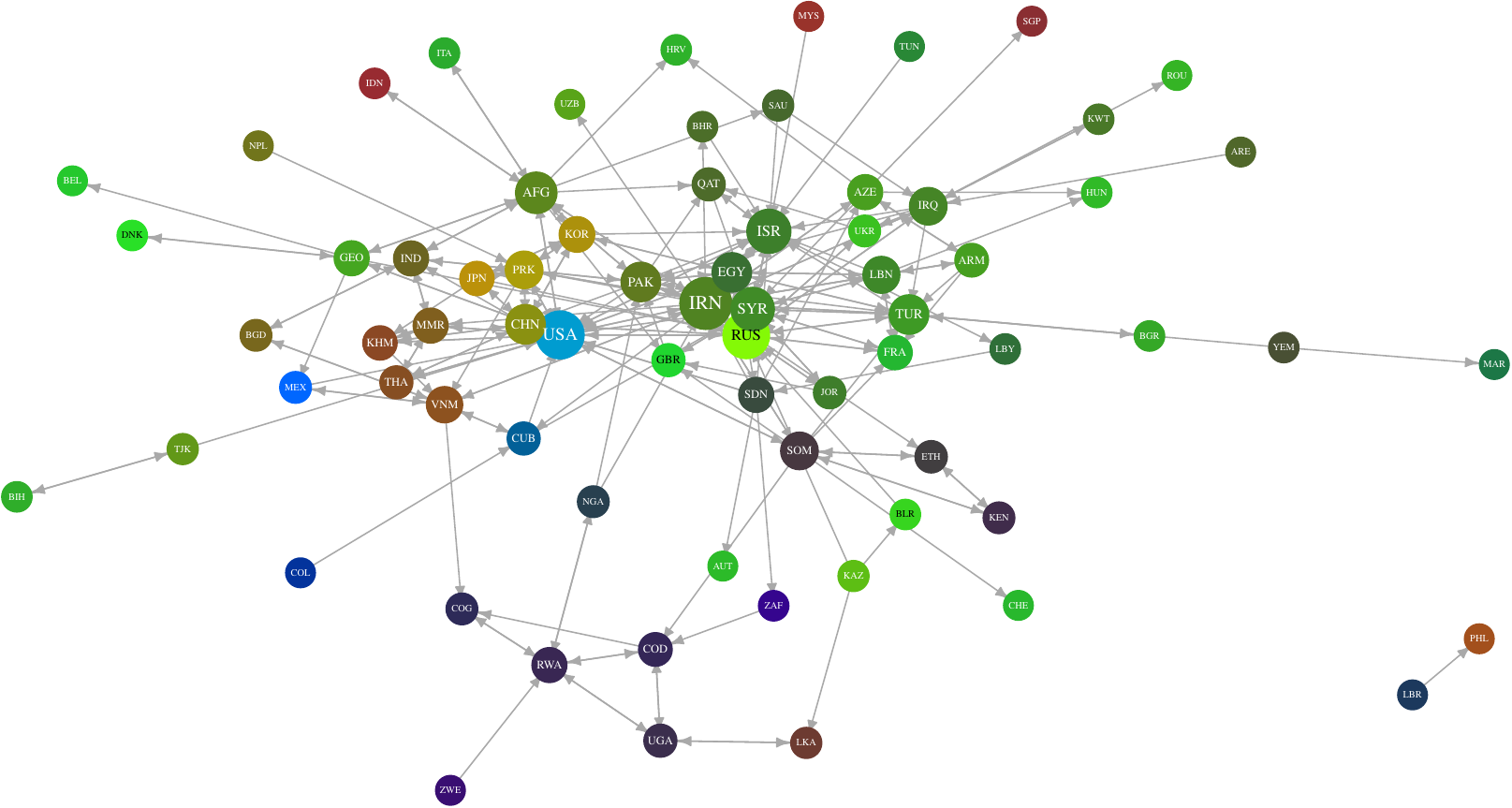} \\
	\end{tabular}
\caption{Influence relationships}
\label{fig:inflRelLong}
\end{figure}
\FloatBarrier




\clearpage
\newpage
\subsection*{Scoring Rules for Count Data}

Scoring rules are penalties $s(y,P)$ introduced with $P$ being the predictive distribution and $y$ the observed value. The goal of researchers interested in prediction is to minimize the expectation of these scores, which is typically calculated by taking the average:

$S = \frac{1}{n} \sum_{i=1}^{n} s(y_{i}, P_{i}),$

where $y_{i}$ refers to the $i^{th}$ observed count and $P_{i}$ the $i^{th}$ predictive distribution. A set of proper scoring rules as defined by \citet{czado:etal:2009} are shown in the list below. For each of these rules, $f(y)$ denotes the predictive probability mass function. $\hat\mu$ and $\hat\sigma$ refer to the mean and standard deviation of the predictive distribution. 

\begin{itemize}
	\item Dawid-Sebastiani score: $s(y,P) = (\frac{y-\hat\mu}{\hat\sigma})^{2} + 2 \times log(\hat\sigma)$
	\item Logarithmic score: $s(y,P) = -log(f(y))$
	\item Brier score:  $s(y,P) = -2f(y) + \sum_{k}f^{2}(k)$ 	
	\item Spherical score: $s(y,P) = \frac{f(y}{\sqrt{\sum_{k}f^{2}(k)}}$
\end{itemize}


\newpage

\section*{\textbf{Funding}}

This work was supported by the National Science Foundation [2017180 to S.M., 1505136 to P.H.].

\section*{\textbf{Acknowledgments}}

We acknowledge the invaluable contributions of our late coauthor, Michael D. Ward, who played a crucial role in the early drafts of this paper. His insights and mentorship were instrumental in shaping the direction of this work.

\clearpage
\bibliography{master}
\bibliographystyle{apsr}
\newpage

\end{document}